\documentclass[10pt,journal,twocolumn]{IEEEtran}
\newif\ifCLASSOPTIONromanappendices \CLASSOPTIONromanappendicestrue
\usepackage{a0size}
\usepackage{amssymb}
\usepackage{multicol}
\usepackage[english]{babel}
\usepackage{epsfig}
\usepackage{bm}
\usepackage{amsfonts,color,amsthm,amsmath, lscape,graphics}
\usepackage{url}
\usepackage{algorithm}
\usepackage{algorithmic}

\newcommand{\tr}{\operatorname{Tr}}
\usepackage[font=footnotesize]{caption}
\usepackage{subcaption}
\usepackage{epstopdf}
\usepackage{cite}
\usepackage{relsize}

\DeclareFontFamily{U}{matha}{\hyphenchar\font45}
\DeclareFontShape{U}{matha}{m}{n}{
      <5> <6> <7> <8> <9> <10> gen * matha
      <10.95> matha10 <12> <14.4> <17.28> <20.74> <24.88> matha12
      }{}
\DeclareSymbolFont{matha}{U}{matha}{m}{n}
\DeclareMathSymbol{\odiv}         {2}{matha}{"63}

\usepackage{hyperref}
\usepackage{fix2col}

\newtheorem{lemma}{Lemma}

\addto\captionsenglish{}

\newcommand{\bh}{\mathbf{h}}

\newcommand{\bv}{\mathbf{v}}
\newcommand{\bV}{\mathbf{V}}

\newcommand{\bU}{\mathbf{U}}
\newcommand{\bI}{\mathbf{I}}

\newcommand{\bd}{\mathbf{d}}

\newcommand{\bC}{\mathbf{C}}
\newcommand{\bD}{\mathbf{D}}
\newcommand{\bH}{\mathbf{H}}

\newcommand{\bG}{\mathbf{G}}
\newcommand{\bp}{\mathbf{p}}

\newcommand{\bQ}{\mathbf{Q}}

\newcommand{\bZ}{\mathbf{Z}}

\renewcommand{\frac}{\dfrac}

\definecolor{myOrange}{rgb}{1,0.5,0}
\definecolor{myGreen}{rgb}{0,0.5,0}
\newcommand{\changeb}[1]{{\color{black}#1}}
\newcommand{\changebb}[1]{{\color{black}#1}}

\begin{document}
\title{Energy-Efficient Flat Precoding for MIMO Systems}

\author{
{Foad~Sohrabi},~\IEEEmembership{Member,~IEEE,}
        Carl~Nuzman,~\IEEEmembership{Member,~IEEE,}\\
	Jinfeng~Du,~\IEEEmembership{Senior~Member,~IEEE,}
        Hong Yang,~\IEEEmembership{Senior~Member,~IEEE}
        and~Harish~Viswanathan,~\IEEEmembership{Fellow,~IEEE}
\thanks{The authors are with are with
Nokia Bell Labs, Murray Hill, NJ 07974, USA (e-mails: \{foad.sohrabi, carl.nuzman, jinfeng.du, h.yang, harish.viswanathan\}@nokia-bell-labs.com). 
}
}
\maketitle
\begin{abstract}
This paper addresses the suboptimal energy efficiency of conventional digital precoding schemes in multiple-input multiple-output (MIMO) systems. Through an analysis of the power amplifier (PA) output power distribution associated with conventional precoders, it is observed that these power distributions can be quite uneven, resulting in large PA backoff (thus low efficiency) and high power consumption. To tackle this issue, we propose a novel approach called flat precoding, which aims to control the flatness of the power distribution within a desired interval. In addition to reducing PA power consumption, flat precoding offers the advantage of requiring smaller saturation levels for PAs, which reduces the size of PAs and lowers the cost. To incorporate the concept of flat power distribution into precoding design, we introduce a new lower-bound per-antenna power constraint alongside the conventional sum power constraint and the upper-bound per-antenna power constraint. By adjusting the lower-bound and upper-bound values, we can effectively control the level of flatness in the power distribution. We then seek to find a flat precoder that satisfies these three sets of constraints while maximizing the weighted sum rate (WSR). In particular, we develop efficient algorithms to design weighted minimum mean squared error (WMMSE) and zero-forcing (ZF)-type precoders with controllable flatness features that maximize WSR. Numerical results demonstrate that complete flat precoding approaches, where the power distribution is a straight line, achieve the best trade-off between spectral efficiency and energy efficiency for existing PA technologies. We also show that the proposed ZF and WMMSE precoding methods can approach the performance of their conventional counterparts with only the sum power constraint, while significantly reducing PA size and power consumption. 

\end{abstract}


\section{Introduction}
Massive multiple-input-multiple-output (MIMO) technology utilizes large antenna arrays to create highly directional beams, offering a promising solution for mitigating pathloss at higher frequencies and meeting the ever-growing demand for spectral efficiency in wireless communication networks \cite{Emil_2020_B5G}. In addition to enabling base stations (BSs) to serve multiple user equipment (UEs) simultaneously on the same time-frequency resources, it also improves transmission energy efficiency by utilizing high beamforming gain \cite{Larsson_2014_CM}. However, the deployment of a large array of antennas can be expensive and power-intensive, particularly when employing traditional fully-digital beamforming techniques. Consequently, incorporating energy efficiency as a crucial performance metric in MIMO system designs has gained significant attention in recent studies \cite{Lopez_2022_CST,Stefan_2023_Mag,Lozano_2023_EE}. The main objective of this paper is to develop new precoding methods that can enhance the energy efficiency and reduce the implementation cost of traditional precoding.

In the literature, there are several conventional precoding schemes that are widely used in practical MIMO systems, e.g., maximum ratio transmission (MRT) \cite{Lo_1999_MRT}, zero-forcing (ZF) \cite{Hong_2013_ZF,Cornelis2023WSRZF}, regularized ZF precoding (RZF) \cite{Peel2005RZF,Emil_RZF_2014}, and weighted minimum mean-square error (WMMSE) \cite{shi2011iteratively,TomLuo2023}. All of these precoding schemes seek to maximize a performance metric, such as sum rate or weighted sum rate (WSR), while satisfying a set of power constraints. One commonly considered power constraint in the literature is the sum power constraint (SPC), e.g., \cite{shi2011iteratively}. Under SPC, the total radiated power of all base station (BS) antennas should not exceed a given power budget. In practice, the maximum permissible radiated power is typically determined based on existing communications regulations, such as the federal communications committee (FCC) regulations on equivalent isotropic radiated power (EIRP). In addition to SPC, the hardware used to implement each antenna branch (i.e., transceiver) can impose further power limitations. For instance, to operate in the linear regime of a power amplifier (PA) in the antenna branch, it is necessary to operate below the saturation power of the PA by a certain margin. To address this, an upper-bound (UB) per-antenna power constraint (PAPC), assuming each antenna is supported by a dedicated 
 PA, has been incorporated in the design of the MIMO precoder, e.g., \cite{TomLuo2023,Wei_2007_PAPC}. This constraint ensures that the power of each antenna branch does not exceed a specified threshold determined based on the hardware characteristics, such as PA characteristics.

However, conventional precoding methods under SPC or UB-PAPC lack a mechanism to control the variations in power across different antennas for the designed precoder. \changebb{In the case of SPC, these power variations ({for a fixed overall radiated power}) can result in higher implementation costs as they necessitate PAs with higher saturation powers. Moreover, the efficiency of PAs typically decreases as operational power decreases. Therefore, for conventional precoding under SPC or UB-PAPC, power variations across antennas mean that some antennas operate at lower powers, leading to higher power/energy consumption for a given radiated power}. In this paper, we present a new concept called \textit{Flat Precoding}, which addresses the issue of power variation across antennas. With flat precoding, it is possible to control the power flatness across antennas within a desired interval. This interval can be designed based on the hardware characteristics, such as PA characteristics. By maintaining power flatness, we can effectively manage overall power consumption and keep the size of PAs at a reasonable level. 
\changebb{Moreover, the proposed flat precoding, which ensures all PAs operate within a tightly controlled output power range, can be effectively leveraged to optimize PA efficiency, achieving peak performance within this targeted operating window.}
A more detailed review of the main contributions of this paper is presented in the following subsection.

\subsection{Main Contributions}
This paper begins by discussing the conceptual advantages of flat precoding, which involves controlling power variations across antennas \changeb{for a given overall radiated power}. We demonstrate that by designing a precoder with complete flatness, it is possible to \changebb{minimize} the required saturation power of the PA in each antenna branch. Consequently, this reduction in PA saturation power results in a smaller PA size and lower implementation cost for the MIMO systems. Furthermore, by recognizing that for the current technologies of PAs, the highest PA efficiency is achieved when operating at/near the saturation level and efficiency decreases as power decreases, we argue that implementing a flat precoding strategy can result in significant power and energy savings which has a great importance for the development of future green communication networks. 

Based on these motivations, we propose modifying the precoding design problem by incorporating a lower-bound (LB) PAPC alongside the conventional UB-PAPC and SPC. By choosing appropriate values for LB-PAPC and UB-PAPC, we can effectively control the level of flatness. By considering WSR as the utility function to maximize, we formulate the multi-user MIMO precoding problem with the new set of LB-PAPCs. To tackle this challenging problem, known to be NP-hard even in the absence of the newly introduced LB-PAPCs \cite{TomLuo2008_NPhard,YaFeng_2011_NPhard}, we employ the WMMSE framework to convert the WSR maximization problem into a WMMSE minimization problem. Similar to WMMSE-type precoders, we utilize the block coordinate descent approach. In block coordinate descent methods like the WMMSE algorithm, a feasible initial point is required to ensure that all solutions in the iterations remain feasible. Therefore, in this paper, we propose a simple method to construct an initial feasible precoder that satisfies the three sets of power constraints, i.e., SPC, UB-PAPC, and LB-PAPC. Further, we demonstrate that for the precoding problem under these three sets of power constraints, the rows of the precoder, the receivers' combining matrices, and the weight matrices that convert the WSR problem into an equivalent WMMSE problem serve as suitable coordinates. By optimizing each coordinate while keeping the others fixed, we obtain an optimal solution for each coordinate. In essence, we illustrate that the idea of flat precoding can be effectively incorporated into the WMMSE algorithm, and call the corresponding method as \textit{Flat-WMMSE}.

While WMMSE-type precoders can achieve excellent performance, simpler approaches such as ZF methods may be more favorable in practice. Therefore, as the next step, this paper extends the concept of flat precoding to ZF precoders. Specifically, we demonstrate that by utilizing the semidefinite relaxation (SDR) technique, which involves matrix lifting and rank-one relaxation, the flat ZF design problem can be approximated as a semi-definite program (SDP). This SDP can be solved using off-the-shelf convex solvers such as CVX \cite{cvx}. Our numerical results indicate that the solution obtained from this WSR flat ZF algorithm is typically rank-one, which aligns with previous observations from various studies \cite{Wisel2008,Ken2017} in the context of multi-user precoding, albeit with different objectives and power constraints.

Although the solution derived from the aforementioned \textit{SDR-Flat ZF} method is nearly optimal, it involves high computational complexity. This issue motivates us to develop a more computationally efficient algorithm for designing a flat ZF precoder. In doing so, we first devise an algorithm for designing the flat ZF precoder by using the primal-dual approach, assuming that the ratio of the received gains for all the layers is given, which we refer to as a fixed-relative-gain (FRG) profile. To complete this \textit{FRG-Flat ZF} method, we subsequently discuss how to obtain the optimal received gain profile that maximizes WSR under only SPC and propose to use this as the target gain profile for the problem of interest in this paper with three types of power constraints.
We note that finding the flat ZF precoder may not be feasible in general. However, as supported by the numerical results, we observe infeasibility of flat ZF precoder only for extreme scenarios where the total number of layers is in the same order as the number of antennas at the BS. However, this is not the main scenario for typical massive MIMO systems, which are the main focus of this paper.

To evaluate the performance of the proposed flat precoding methods, we conduct extensive simulations for various communication settings and make the following observations:
\begin{itemize}
\item The proposed flat precoding methods effectively control the level of power flatness across antennas within a desired interval, which can be optimized based on hardware characteristics such as PAs.
\item It is demonstrated that imposing flatness, even complete flatness, only leads to marginal degradation in spectral efficiency, particularly in massive MIMO scenarios.
\item For the considered PAs, specifically Doherty Gallium Nitride (GaN) PAs \cite{Doherty2016}, in which the PA efficiency decays exponentially as the gap \changeb{(in dB)} to saturation increases, it is shown that completely flat precoding significantly reduces power consumption and the required PA saturation levels. Therefore, in combination with the previous observation of almost unchanged spectral efficiency, it is concluded that completely flat precoding is the most appropriate approach given the current PA technology.
\item The proposed SDR-Flat ZF approach which utilizes an SDP solver yields excellent performance, but its computational complexity and CPU time requirements are quite high. However, the proposed  FRG-Flat ZF algorithm achieves nearly the same performance with processing times several orders of magnitude faster.
\item While the proposed fixed-ratio FRG-Flat ZF algorithm is the most efficient among the proposed flat precoding methods, it is shown that the Flat WMMSE precoders offer certain advantages over ZF solutions. For instance, when the number of antennas is comparable to the number of UEs or in multi-layer MIMO scenarios.
\end{itemize}

\subsection{Related Works} 

Conventional communication system design primarily focuses on optimizing the trade-off between performance and computational complexity. For example, this can involve maximizing the throughput while keeping the computational complexity within reasonable limits. In the context of MIMO precoding, which is the main focus of this paper, the conventional approach typically aims to maximize a measure of spectral efficiency, such as sum rate \cite{Foad2016Hybrid}, weighted sum rate \cite{shi2011iteratively}, or min rate \cite{Manijeh2019maxmin}, while respecting regulatory and hardware constraints.

With this mindset, various precoding methods have been developed for multi-user MIMO systems. One of the simplest forms of linear precoding is MRT precoding \cite{Lo_1999_MRT}, where the precoder is designed to maximize the power of the intended signals at the receivers. While MRT algorithm is quite simple and computationally friendly, it is unable to handle inter-layer or inter-user interference. As a result, simple yet effective ZF type precoders, e.g., \cite{Hong_2013_ZF,Cornelis2023WSRZF}, are developed, in which interference under perfect channel state information can be completely canceled. However, in scenarios where the number of spatial degrees of freedom is insufficient or the channels are ill-conditioned due to high correlation, the ZF solution performs poorly as it aims to ensure complete interference nulling \cite{Peel2005RZF}. To address this drawback of ZF, a regularization term is introduced in the RZF precoding method. The regularization term is chosen to make a balance between interference mitigation and power enhancement \cite{Emil_RZF_2014,Peel2005RZF}. Another alternative approach that has attracted significant attention is WMMSE precoding \cite{shi2011iteratively}. In this method, the WSR problem is converted into a weighted MSE minimization problem, and a block coordinate descent method is employed to find the precoder. The iterative WMMSE methods are guaranteed to converge to a stationary solution, and numerous examples have reported their near-optimal performance \cite{Juseong2024AnalogFeed}.

Most of the aforementioned precoding methods were initially developed under SPC, which restricts the total power of all BS antennas to a given budget. This power budget is typically determined based on regulations governing the maximum EIRP which happens in the broadside direction \cite{Kim2012Globe}. However, in practice, each antenna branch, consisting of a PA, has its own individual power budget \cite{Zheng2007PAPC,Victor2019PA}. As a result, subsequent research has aimed to ensure that the designed precoders also satisfy UB-PAPC. In its simplest form, to satisfy UB-PAPC, we can scale down the precoder designed by the SPC precoding schemes \cite{Lee2013Scale}. However, this approach leaves a significant portion of the power budget unused at some antennas, resulting in potential performance degradation. This motivates the extension of the precoding designs under SPC to incorporate UB-PAPC, where the UB-PAPC is set based on the maximum power budget of each antenna branch.

For instance, some works (e.g., \cite{Wisel2008, Rui2010, Pham2018}) have focused on developing variants of ZF precoders under UB-PAPCs for MIMO systems. In particular, \cite{Wisel2008} demonstrates that by applying SDR, the ZF problem can be formulated as a convex problem and optimally solved. The authors of \cite{Wisel2008} further prove that the relaxed problem always has a rank-one solution and propose a method to recover the rank-one solution. In this paper, we show that it is possible to use the same SDR technique to find a flat ZF precoder. However, unlike \cite{Wisel2008} which only considers UB-PAPC, there is no guarantee that the solution of the relaxed problem is rank one. Nevertheless, in practice, we observe that in most cases, the solution is rank one. For the remaining cases, we propose a technique to make it feasible for the power constraints, although the ZF condition may no longer be imposed.

\changeb{While it is possible to employ convex optimization tools and packages to find a ZF precoder under PAPC constraints}, the practical computational complexity of such approaches may still be high. To reduce the computational complexity of finding the desired ZF solution, the subgradient and approximation methods have been proposed in the literature, e.g., \cite{Rui2010, Pham2018}. In this paper, instead of following these strategies, we propose a highly efficient iterative method called FRG-Flat ZF by exploiting the primal-dual approach. We show that the performance of FRG-Flat ZF can approach that of the more relaxed baselines only under SPC.

Given the urgent threat of climate change, it becomes increasingly important to minimize energy usage achieve sustainability and support future technologies \cite{Stefan_2023_Mag}. In the wireless communications industry, the development of greener design in physical layer technologies is crucial for sustainable growth in 6G networks and beyond \cite{Han2021Green}. This paper argues that the traditional approach of optimizing performance and computational complexity alone is insufficient for designing future wireless communication systems. Recognizing the need for sustainability, recent research has focused on energy efficiency as a third dimension for system optimization \cite{Lopez_2022_CST,Lozano_2023_EE}. This paper proposes the  concept of flat precoding as a method to reduce energy consumption and implementation costs.

\subsection{Paper Organization and Notations}

The remainder of the paper is organized as follows. Section~\ref{sec:sys} describes the system model and the WSR maximization problem formulation. Section~\ref{sec:motiv} highlights the motivation behind designing a digital precoder that controls power variations across antennas. Section~\ref{sec:flat_WMMSE_CF} proposes a WMMSE precoding method with controllable flatness feature. Section~\ref{sec:flat_ZF_CF} extends the idea   of flat precoding to the ZF-type precoders. Section~\ref{sec:flat_ZF_CF_SO} develops a more computationally efficient algorithm by fixing the received relative gain profile. Section~\ref{sec:numerical} presents simulation results. Finally, conclusions are drawn in Section~\ref{sec:conc}.

This paper uses lower-case letters for scalars, lower-case bold-face letters for vectors, and upper-case bold-face letters for matrices. The element in the $i$-th row and the $j$-th column of $\bV$ is denoted by $[\bV]_{i,j}$; \changeb{$\operatorname{Diag}(\bv)$} returns a diagonal matrix with elements from $\bv$; \changeb{$\operatorname{diag}(\bV)$ returns a vector containing the diagonal elements from $\bV$}; $\operatorname{blkdiag}(\{\bV_k\}_{k=1}^{K})$ returns a block diagonal matrix with $\bV_k$'s as the diagonal blocks; \changeb{$\max(\bv, x)$ and $\min(\bv, x)$ return vectors where each element is the maximum or minimum, respectively, of the corresponding element in the vector $\bv$ and the scalar $x$}; and $(x)^{+}$ represents $\max(x,0)$. Superscripts $(\cdot)^T$, $(\cdot)^{H}$, and $(\cdot)^{-1}$ denote the transpose, Hermitian transpose, and inverse of a matrix, respectively. The $N\times N$ identity matrix is denoted by $\mathbf{I}_N$; the all-ones vector with appropriate dimension is denoted by $\mathbf{1}$; $\mathbb{C}^{m\times n}$ represents an $m$ by $n$ dimensional complex space; and $\mathcal{CN}(\mathbf{0},\mathbf{R})$ represents the zero-mean circularly symmetric complex Gaussian distribution with a covariance matrix $\mathbf{R}$. Notations \changeb{$\operatorname{log}(\cdot)$}, $\operatorname{log}_{10}(\cdot)$, $\operatorname{Tr}(\cdot)$, and $\lceil \cdot \rceil$ represent the \changeb{natural} logarithm, decimal logarithm, trace, and ceiling operators, respectively. Finally, $|\cdot|_2$ indicates the Euclidean vector norm; $|\cdot|$ represents the determinant or absolute value depending on the context; and $\odiv$ denotes the element-wise division. 
\section{System Model and Problem Formulation}
\label{sec:sys}

 Consider a downlink multi-user MIMO system in which a BS with $N$ antennas serves $K$ UEs each with $M$ antennas. Let $\mathbf{s}_k \in \mathbb{C}^{L}$ denote the vector of intended symbols for UE $k$, and $\mathbf{D}_k \in \mathbb{C}^{N\times L}$ denote the digital linear precoder associated with UE $k$ {\footnote{\changeb{While this paper focuses on fully-digital precoding, the proposed algorithms can be readily applied to digital precoding design in practical partially-connected hybrid structures. Extending them to fully-connected structures is an interesting area for future research.}}}. Then, the transmitted signal can be written as:
\begin{equation}
\mathbf{x} = \sum_{k=1}^{K} \mathbf{D}_k \mathbf{s}_k.
\end{equation}

Assuming flat-fading narrow-band channel model, the received signal at UE $k$ can be expressed as:
\begin{eqnarray}
\mathbf{y}_k &=& \mathbf{H}_k \mathbf{x} + \mathbf{n}_k\\ \nonumber
&=&  \mathbf{H}_k \mathbf{D}_k \mathbf{s}_k + \sum_{j\not=k} \mathbf{H}_k \mathbf{D}_j \mathbf{s}_j + \mathbf{n}_k,
\end{eqnarray}      
where $\mathbf{H}_k \in \mathbb{C}^{M \times N}$ is the channel matrix from the BS to the $k$-th UE and $\mathbf{n}_k \in \mathbb{C}^{M}$ is the additive white Gaussian noise vector with distribution $\mathcal{CN}\left(\mathbf{0},\sigma^2\mathbf{I}_{M}\right)$. 
In such a system, the following inequalities hold between different system dimensions: $N \geq K M \geq  KL$.

Furthermore, under Gaussian signaling assumption, i.e., $\mathbf{s}_k \sim \mathcal{CN}(\mathbf{0},\mathbf{I}_{L})$, the achievable rate for UE $k$ is given by:
\begin{equation}
\label{eq:sUE_rate}
\begin{aligned}
R_k\triangleq \changeb{\log} &\left| \mathbf{I}_{M}+\mathbf{H}_{k} \mathbf{D}_{k} \mathbf{D}_{k}^{H} \mathbf{H}_{k}^{H}  \bC_k^{-1}\right|, 
\end{aligned}
\end{equation}
where $\bC_k = \sum_{j \neq k} \mathbf{H}_{k} \mathbf{D}_{j} \mathbf{D}_{j}^{H} \mathbf{H}_{k}^{H}+\sigma^{2} \mathbf{I}_{M}$.

In this paper, we are interested to maximize the spectral efficiency of the system. Given the weight priorities for the UEs $\{\alpha_k\}_{k=1}^{K}$, the spectral efficiency is modeled as WSR:
\begin{equation}
\begin{aligned}
R({{\mathbf{D}}})=\sum_{k=1}^{K} \alpha_{k} R_k. 
\end{aligned}
\end{equation}

The problem is how to design the overall digital precoder ${{\mathbf{D}}}\triangleq\left[ \mathbf{D}_{1},\ldots, \mathbf{D}_{K}\right]\in \mathbb{C}^{N \times KL }$ in order to maximize WSR, i.e., $R(\mathbf{D})$, while satisfying the required power constraints as:
 \begin{equation}
\begin{aligned}
\max_{{{\mathbf{D}}} \in \mathcal{P}} ~ R({{\mathbf{D}}}), 
\end{aligned}
\end{equation}
where $\mathcal{P}$ denotes the set of feasible precoders that satisfy the required power constraints. In the next section, we first overview the conventional power constraints. Subsequently, we introduce a new set of power constraints which enables us to control the power variation across antennas, aiming to enhance the energy efficiency of the system.

\section{Motivation for Controlling Flatness of Digital Precoder}
\label{sec:motiv}
In the existing literature, two main categories of power constraints have conventionally been explored. The first category is known as the sum power constraint, i.e., SPC. The reason for implementing SPC is to control the amount of the average transmitted power from the BS, thereby restricting the potential interference it could introduce to neighboring cells or networks. If $P_\text{TX}$ denotes the maximum permissible average radiated power, SPC is typically defined as follows:
\begin{equation}
\label{eq:SPC}
\begin{aligned}
\sum_{k=1}^{K} \operatorname{Tr} \left( \mathbf{D}_k \mathbf{D}_k^H \right)   \leq P_\text{TX}.
\end{aligned}
\end{equation}

\begin{figure*}[t]
  \centering
  \begin{subfigure}{2.63in}
    \includegraphics[width=\textwidth]{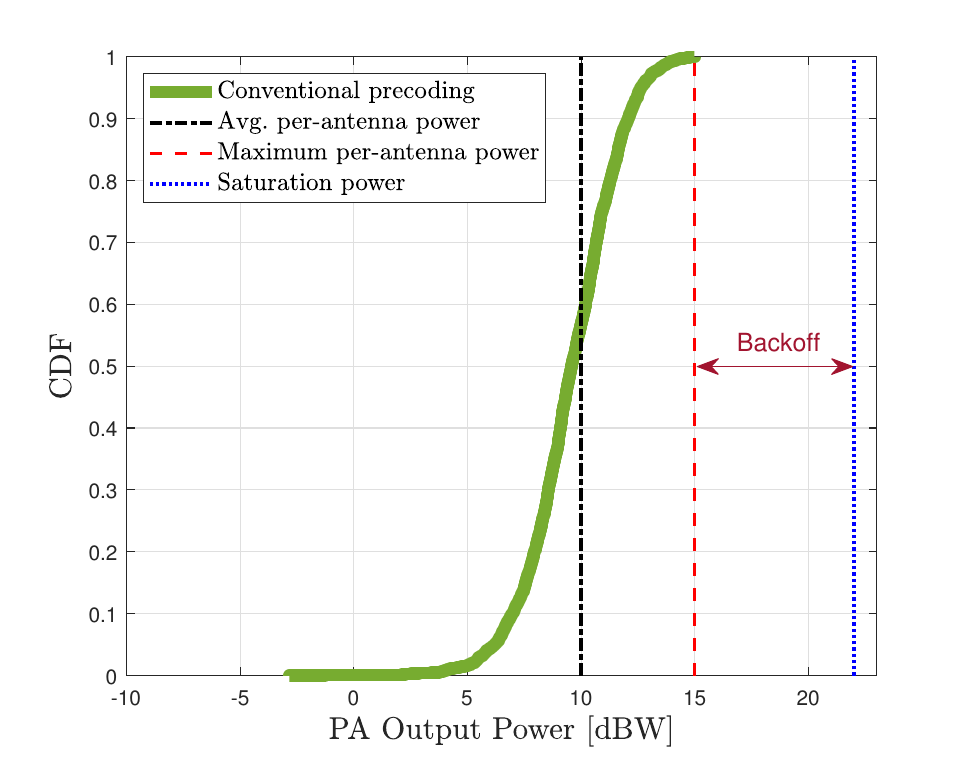}
    \caption{The conventional precoding schemes.}
    \label{subfig1:conv}
  \end{subfigure}%
  \hfil
  \begin{subfigure}{2.63in}
    \includegraphics[width=\textwidth]{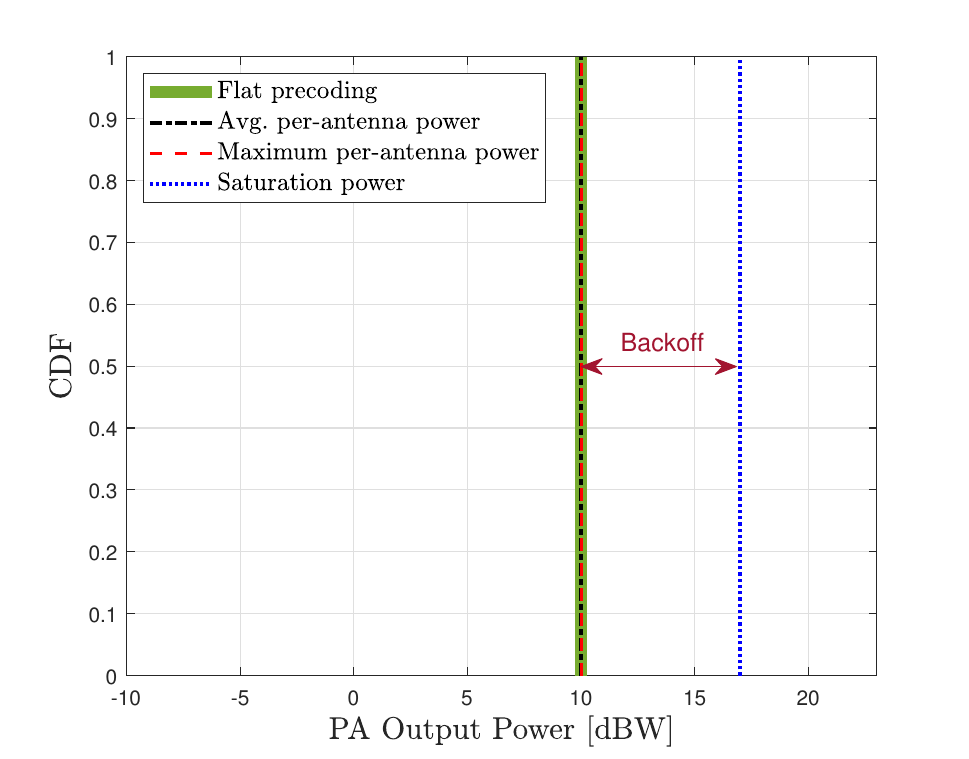}
    \caption{The proposed completely flat precoding scheme.}
    \label{subfig1:flat}
  \end{subfigure}
  \caption{The cumulative distribution function (CDF) curve of the PA output power distribution. \changeb{It should be noted that the CDF curve for the conventional precoding method is derived from running a ZF precoder in a system with $K=8$ single-antenna users served by a BS with $16$ antennas in a Rayleigh fading channel environment. The general behavior of conventional precoding methods under different parameter settings is observed to follow the same trend, characterized by a wide PA output power distribution. }}
  \label{fig1:overall}
\end{figure*}

The second category of conventional power constraints involves imposing an UB limit on the average transmitted power for each antenna. This kind of constraint, known as per-antenna power constraint, i.e., PAPC, originates from the need to restrict the output power of the PA connected to each antenna ensuring that the PA operates in its linear regime. To meet such power limits of the PAs, the designed digital precoder must satisfy the following UB-PAPC:
\begin{equation}
\label{eq:UB-PAPC}
\begin{aligned}
\sum_{k=1}^{K} \operatorname{diag} \left( \mathbf{D}_k \mathbf{D}_k^H \right)  \leq \changeb{P^\text{ub}_\text{PAPC} {\cdot} \mathbf{1}_{N},}
\end{aligned}
\end{equation}
where $P^\text{ub}_\text{PAPC}$ is conventionally calculated based on the PA power limit\footnote{In this paper, we consider a common PAPC constraint across all antennas. However, the discussions can be easily extended to the case with different PAPC constraints for different antennas.}. We note that the conventional SPC in \eqref{eq:SPC} and UB-PAPC in \eqref{eq:UB-PAPC} stem from distinct regulations and requirements. Consequently, there is no specific relationship between $P_\text{TX}$ and $P^\text{ub}_\text{PAPC}$. 

\changebb{It is worth mentioning that the SPC in \eqref{eq:SPC} is typically active (satisfied with equality) for the optimal precoder because maximizing the spectral efficiency requires full utilization of the available power (if possible). For the precoding design only under SPC, it is well known that the SPC should be satisfied by equality at the optimal solution \cite{shi2011iteratively}. However, for the scenario where both SPC and PAPC-type constraints are present, the optimal solution becomes more intricate because all these constraints must be jointly satisfied. Nevertheless, as long as the PAPC-type constraints are chosen appropriately, the optimized solutions will typically converge to use the entire $P_\text{TX}$. Therefore, in the discussion within this section, we assume that the overall radiated power remains fixed. However, it is important to emphasize that this assumption is not enforced in our problem formulation, nor in the proposed algorithms presented in the subsequent sections.}

This paper recognizes that using conventional precoding schemes solely based on these two types of power constraints may result in significant power variations across different antennas (see Fig.~\ref{subfig1:conv}). This, combined with the fact that the power efficiency of a PA connected to an antenna depends on the operational power of that antenna, can result in a decline in power efficiency of the system. This decline can be particularly substantial, especially in massive MIMO systems with a large number of PAs. 
For instance, in current technology, the efficiency of the PA decreases as the operational PA output power reduces and moves further away from the saturation level \cite{Doherty2016}. Hence, employing conventional precoding schemes with a wide average power distribution, as illustrated in Fig.~\ref{fig1:overall}, results in a high probability of a PA operating at notably low efficiency. \changebb{This low efficiency of some of the PAs leads to increased overall power consumption, compared with a scenario with the same total overall radiated power divided equally among PAs.}\footnote{\changebb{We emphasize that the power consumption of a PA operating at lower output power (and thus lower efficiency) does not necessarily increase. However, when considering the collective behavior of PAs with a fixed overall radiated power, a narrower dynamic range can lead to overall power savings due to the ability to operate all PAs at higher efficiencies.}}

To address this issue, we propose to control the power variation (or equivalently flatness) across antennas by properly designing the precoder. In particular, we propose to modify the precoder design problem by adding a new LB constraint $P^\text{lb}_\text{PAPC}$ on the power of each antenna as:
\begin{equation}
\begin{aligned}
\sum_{k=1}^{K} \operatorname{diag} \left( \mathbf{D}_k \mathbf{D}_k^H \right)  \geq \changeb{P^\text{lb}_\text{PAPC} \cdot \mathbf{1}_{N}}.
\end{aligned}
\end{equation}
 By appropriately setting the value of $P^\text{lb}_\text{PAPC}$ and $P^\text{ub}_\text{PAPC}$, we can essentially control the level of flatness of the PA output power distribution. 
For example, $P^\text{lb}_\text{PAPC}$ and $P^\text{ub}_\text{PAPC}$ can be set based on PA characteristics to ensure that the PA efficiency is maintained at a specific level. In the extreme case, by setting $P^\text{lb}_\text{PAPC}=P^\text{ub}_\text{PAPC}$, we can achieve a completely \textit{flat precoding} structure that ensures that there are no power variations over different antennas. For the current technology of PAs, employing such complete flat precoding implies that all PAs consistently operate at their maximum possible efficiency. \changeb{Fig.~\ref{subfig1:flat}} illustrates an example for the PA output power distribution of a completely flat precoder.

The other advantage of employing flat precoding is from the implementation cost perspective. By comparing the PA output power distribution of the conventional precoding and the flat precoding, respectively in Fig.~\ref{subfig1:conv} and \changeb{Fig.~\ref{subfig1:flat}}, we can see that for a given average per-antenna radiated power and a fixed power back-off from the saturation power (needed to account for signal/waveform power variations), the required saturation level for flat precoding can be extremely lower than that of the conventional precoding (e.g., in Fig.~1., the saturation power reduction is about $5$dBW.). This is because the saturation level is mainly determined by the peak radiated power, and employing  flat precoding reduces the required peak power. Thus, the implementation of flat precoding requires smaller PAs, offering a potential reduction in system implementation cost. This cost-effectiveness is crucial for ensuring the scalability and sustainability of future extreme MIMO systems.

From the above discussion, it becomes apparent that the proposed flat precoding idea has a great potential to significantly improve the power and cost efficiency of the system. In the remaining sections, our main objective is to devise efficient algorithms for designing near-optimal flat precoding to maximize spectral efficiency performance. We will then present numerical evidence demonstrating that the proposed flat precoding methods can achieve spectral efficiency levels comparable to conventional methods, thereby  exhibiting superior energy efficiency due to their enhanced power efficiency.

\section{{WMMSE Precoding with Controllable Flatness}}
\label{sec:flat_WMMSE_CF}
In this section, we seek to develop a generic algorithm to tackle the following precoding design problem with three types of power constraints (i.e., SPC, UB-PAPC, and LB-PAPC):
\begin{subequations} \label{eq:WMMSE_problem}
\begin{eqnarray} 
\max_{\{\mathbf{D}_k\}_{\forall k}} &~~& \sum_{k=1}^{K} \alpha_{k} R_k \\ 
\text{s.t.~~} &~~& \sum_{k=1}^{K} \operatorname{Tr}\left( \mathbf{D}_k \mathbf{D}_k^H \right)   \leq P_\text{TX},  \\ 
~~~ &~~&  \sum_{k=1}^{K} \operatorname{diag} \left( \mathbf{D}_k \mathbf{D}_k^H \right)  \leq \changeb{P^\text{ub}_\text{PAPC} \cdot \mathbf{1}_{N}},\\  
~~~ &~~& \sum_{k=1}^{K} \operatorname{diag} \left( \mathbf{D}_k \mathbf{D}_k^H \right)  \geq \changeb{P^\text{lb}_\text{PAPC} \cdot \mathbf{1}_{N}}.
\end{eqnarray}
\end{subequations}
\changeb{In this formulation, we assume that the power limits including $P^\text{ub}_\text{PAPC}$ and $P^\text{lb}_\text{PAPC}$ are given and satisfy $N P^\text{lb}_\text{PAPC} \leq P_\text{TX}$ and $P^\text{lb}_\text{PAPC} \leq P^\text{ub}_\text{PAPC}$ to ensure the feasibility of the problem.} In Section~\ref{subsec:power_consumption}, we will provide an example approach for determining $P^\text{ub}_\text{PAPC}$ and $P^\text{lb}_\text{PAPC}$.

The WSR maximization problem \eqref{eq:WMMSE_problem} is a challenging non-convex problem. One customary practice to tackle such WSR maximization problems is to convert them into their equivalent weighted MSE minimization problems. To do so, we can use the following lemma \cite{Shi2015}.

\begin{lemma}\label{lemma_WMMSE}
Given $\mathbf{B}\in\mathbb{C}^{n\times l}$
and any positive definite matrix $\mathbf{N}\in\mathbb{C}^{n\times n}$, the following equation holds:
\begin{equation}\label{eq_WMMSE_nature}
\begin{aligned}
~~~~~\changeb{\log} & \left| \mathbf{I}_n+\mathbf{B}\mathbf{B}^H\mathbf{N}^{-1} \right| \\ &= \max_{\mathbf{\Omega}\succ \mathbf{0},\mathbf{\Psi}} \changeb{\log} \left|\mathbf{\Omega}\right| -\tr\left(\mathbf{\Omega}\mathbf{E}\left(\mathbf{\Psi},\mathbf{B}\right)\right)+l,
\end{aligned}
\end{equation}
where $\mathbf{\Psi}\in\mathbb{C}^{n\times l}$ and $\mathbf{\Omega}\in\mathbb{C}^{l\times l}$ are auxiliary variables, and
$\mathbf{E}\left(\mathbf{\Psi},\mathbf{B}\right) \triangleq   \left(\mathbf{I}_l -\mathbf{\Psi}^H\mathbf{B}\right)\left(\mathbf{I}_l-\mathbf{\Psi}^H\mathbf{B}\right)^H + \mathbf{\Psi}^H\mathbf{N}\mathbf{\Psi}
$.
Furthermore, the optimal solutions of $\mathbf{\Psi}$ and $\mathbf{\Omega}$ for the right-hand side of \eqref{eq_WMMSE_nature} are respectively given by
\begin{subequations}
\begin{eqnarray}\label{eq_lemma_optimal_U}
{\mathbf{\Psi}}^\star&=&\left(\mathbf{N}+\mathbf{B}\mathbf{B}^H\right)^{-1}\mathbf{B},\\
{\mathbf{\Omega}}^\star&=&\left(\mathbf{E}\left({\mathbf{\Psi}^\star},\mathbf{B}\right)\right)^{-1}=\left(\mathbf{I}_l -{{\mathbf{\Psi}}^\star}^H\mathbf{B}\right)^{-1}.
\label{eq_lemma_optimal_W}
\end{eqnarray}
\end{subequations}

\end{lemma}

We employ Lemma~\ref{lemma_WMMSE} to transform the WSR maximization in \eqref{eq:WMMSE_problem} to a weighted MSE minimization problem. We note that analogous steps were taken in \cite{TomLuo2023} for a WSR maximization problem, albeit with either SPC or UB-PAPC. Here, we undertake this transformation while simultaneously satisfying three sets of constraints, i.e., SPC, UB-PAPC, and LB-PAPC. 

\changeb{Let us define 
$\mathbf{B}_k\triangleq\mathbf{H}_k\mathbf{D}_k$ and $\mathbf{N}_k\triangleq \bC_k$.} Further, let $\mathcal{P}$ denote the feasibility set of the digital precoder that satisfies all three types of power constraints in problem \eqref{eq:WMMSE_problem}. By these considerations, based on Lemma~\ref{lemma_WMMSE}, we can rewrite problem \eqref{eq:WMMSE_problem} as the following MSE minimization problem:
\begin{equation}\label{eq_transform_problem_WMMSE}
\begin{aligned}
\min_{\mathbf{W},\mathbf{U},\mathbf{D}}\  \ & \sum_{k=1}^{K} \alpha_{k} \left(\operatorname{Tr}\left(\mathbf{W}_{k}\mathbf{E}_{k}\right) - \changeb{\log} \left| \mathbf{W}_{k}\right| \right) \\
\text { s.t.~~}\  \ & \mathbf{D} \in \mathcal{P}, 
\end{aligned}
\end{equation}
where $\mathbf{W}\triangleq\left\{\mathbf{W}_{k}\right\}_{k=1}^{K}$ and $\mathbf{U}\triangleq\left\{\mathbf{U}_{k}\right\}_{k=1}^{K}$ are two auxiliary variables (respectively playing the same roles as $\mathbf{\Omega}$ and $\mathbf{\Psi}$ in Lemma \ref{lemma_WMMSE}), and
\begin{equation}\label{MSE_matrix_WMMSE}
\begin{aligned}
\mathbf{E}_{k}  \triangleq &(\mathbf{I}_L -\mathbf{U}_{k}^H \mathbf{H}_{k} \mathbf{D}_{k})(\mathbf{I}_L - \mathbf{U}_{k}^H \mathbf{H}_{k}\mathbf{D}_{k})^H  + \mathbf{U}_{k}^H \bC_k \mathbf{U}_{k}.
\end{aligned}
\end{equation}
Here, $\mathbf{U}_k$ can be interpreted as a combiner used at UE $k$ to recover its intended symbols, $\mathbf{E}_k$ as the corresponding MSE, and $\mathbf{W}_k$ as the appropriate MSE weights, ensuring the equivalence of the WSR maximization problem \eqref{eq:WMMSE_problem} and the weighted MSE minimization problem \eqref{eq_transform_problem_WMMSE}.

In this section, we propose an algorithm to tackle the MSE minimization problem in \eqref{eq_transform_problem_WMMSE}. Since, a digital precoder obtained by solving problem \eqref{eq_transform_problem_WMMSE} satisfies both LB-PAPCs and UB-PAPCs (and the difference between these two bounds can be made arbitrarily small), we call the proposed method as the {\it{Flat} WMMSE} precoding algorithm.

The MSE minimization problem in \eqref{eq_transform_problem_WMMSE} is still a non-convex optimization problem in terms of all the optimization variables. However, we can decompose the problem in terms of a set of variables (called block variables) such that given all block variables except only one variable, the remaining problem is convex in terms of that variable. For the considered problem of interest, the natural choice for such variables are ${\mathbf{U}}$, $\mathbf{W}$, and rows of ${{\mathbf{D}}}$ denoted by $\bd^H_n, \forall n = 1,\ldots,N$.

Based on Lemma~\ref{lemma_WMMSE} (i.e., \eqref{eq_lemma_optimal_U}), we can readily find a closed-form expression for $\mathbf{U}=\left\{\mathbf{U}_{k}\right\}_{k=1}^{K}$ given ${{\mathbf{D}}}$ and $\mathbf{W}$:
\begin{equation}\label{U_update}
\begin{aligned}
\mathbf{U}_{k} = \left(\displaystyle{\sum}_{j=1}^K \mathbf{H}_{k}\mathbf{D}_{j}\mathbf{D}_{j}^{H}\mathbf{H}_{k}^{H}+\sigma^{2} \mathbf{I}_M \right)^{-1}\mathbf{H}_{k}\mathbf{D}_{k},\; \forall k.
\end{aligned}
\end{equation}
 Furthermore, based on \eqref{eq_lemma_optimal_W}, the optimal $\mathbf{W}=\left\{\mathbf{W}_{k}\right\}_{k=1}^{K}$ for given ${\mathbf{U}}$ and ${\mathbf{D}}$ can be obtained by:
\begin{equation}\label{W_update}
\begin{aligned}
\mathbf{W}_{k} = \left(\mathbf{I}_L-\mathbf{U}_k^H\mathbf{H}_k\mathbf{D}_k \right)^{-1},\; \forall k.
\end{aligned}
\end{equation}

The only remaining question to address is how to find the optimal design for a row of ${{\mathbf{D}}}$ (denoted by $\bd^H_n$) when all other rows, ${\mathbf{U}}$, and ${\mathbf{W}}$ are fixed. 
Following the same steps as in \cite{TomLuo2023}, we can show that the problem of designing ${\mathbf{d}}_n$ given all other block variables can be written as:
\begin{subequations}\label{P_PAPC_final}
    \begin{eqnarray}
        \min_{{\mathbf{d}}_n} &
        &a_{n} \|{\mathbf{d}}_n\|_2^{2}+2\Re e\left(\mathbf{b}_{n}^H{\mathbf{d}}_n
        \right)
        \\
        \text{s.t.} &    &  \|{\mathbf{d}}_n\|_2^2\leq P^\text{res}_n ,\\
	  &&  \|{\mathbf{d}}_n\|_2^2 \leq P^\text{ub}_\text{PAPC}, \\
        &&  \|{\mathbf{d}}_n\|_2^2 \geq P^\text{lb}_\text{PAPC},
    \end{eqnarray}
\end{subequations}
where 
\begin{subequations}\label{eq_a_m}
\begin{eqnarray}
a_{n}&\triangleq&{\mathbf{h}}_{n}^{H}\mathbf{A} {\mathbf{h}}_{n},\\
\mathbf{b}_{n} &\triangleq& -\mathbf{B}{\mathbf{h}}_{n}+\displaystyle{\sum}_{\ell \neq n}{\mathbf{d}}_\ell {\mathbf{h}}_{\ell}^{H}\mathbf{A} {\mathbf{h}}_{n},\\
\mathbf{A}&\triangleq&\operatorname{blkdiag}\left( \left\{ \alpha_{k}\mathbf{U}_{k}\mathbf{W}_{k}\mathbf{U}_{k}^{H} \right\}_{k=1}^{K} \right),\\
\mathbf{B}&\triangleq&\operatorname{blkdiag}\left( \left\{ \alpha_{k}\mathbf{W}_{k}\mathbf{U}_{k}^{H} \right\}_{k=1}^{K} \right),\\
P^\text{res}_n &\triangleq& P_\text{TX} - \displaystyle{\sum}_{\ell\not =n} \|{\mathbf{d}}_\ell\|_2^2,
\end{eqnarray}
\end{subequations}
and ${\mathbf{h}}_{n}$ being the $n$-th column of the overall channel matrix $\mathbf{H} \triangleq \left[\mathbf{H}_1^T, \ldots, \mathbf{H}_K^T \right]^T$. By looking at the quadratic objective function of \eqref{P_PAPC_final}, we can see that the optimal design of ${{\mathbf{d}}_n}$ should be in the direction of $-\mathbf{b}_n$. Ignoring the constraints, we can show that the maximizer of \eqref{P_PAPC_final} is given by $\tfrac{-1}{a_n}\mathbf{b}_n$. But for the actual problem with constraints, the scaling factor for $-\mathbf{b}_n$ should be designed such that it ensures all the constraints are satisfied, i.e.,
\begin{equation}\label{P_PAPC_solution_brief}
{\mathbf{d}}_{n}=-\mathbf{b}_{n} \changeb{\cdot} ~ {\max} \left( \min \left( \frac{1}{a_{n}},\frac{\sqrt{P^\text{ub}_n}}{ \|\mathbf{b}_{n}\|_2} \right), {{\frac{\sqrt{P^\text{lb}_\text{PAPC}}}{ \|\mathbf{b}_{n}\|_2}}} \right),
\end{equation}
where $P^\text{ub}_n \triangleq \min\left(P^\text{res}_n, P^\text{ub}_\text{PAPC}\right)$.

The summary of the Flat WMMSE precoding algorithm is illustrated in Algorithm \ref{alg_PAPC-WMMSE}. This algorithm starts from a (feasible) initial precoder and iteratively updates $\mathbf{U}$, $\mathbf{W}$, and ${\mathbf{d}}_{n}, \forall n$, until convergence or reaching to the predefined maximum number of iterations. We emphasize that such a block coordinate descent algorithm started from an initial feasible solution is guaranteed to converge to a stationary solution of the original WSR maximization problem \cite{TomLuo2023}. 

\begin{algorithm}[!tb]
    \renewcommand{\algorithmicrequire}{\textbf{Input:}} 
    \renewcommand{\algorithmicensure}{\textbf{Output:}}
    \caption{Flat WMMSE Precoding}  \label{alg_PAPC-WMMSE}
    \begin{algorithmic}[1]
    \REQUIRE System dimensions $N$, $K$, $M$, and $L$; channel matrix $\mathbf{H} = [\bh_1,\ldots,\bh_N]$; priority weights $\{\alpha_k\}_{\forall k}$; noise variance $\sigma^2$;  power constraints $P_\text{TX}$, $P^\text{ub}_\text{PAPC}$, and $P^\text{lb}_\text{PAPC}$; convergence threshold $\epsilon$; maximum number of iterations $i_\text{max}$.
    \STATE Initialize \textbf{(feasible)} digital precoder $\mathbf{D} = \left[ \mathbf{D}_{1},\ldots, \mathbf{D}_{K}\right]$, $\mathbf{W}_k=\mathbf{I},\forall k$, and $i=1$.
    \REPEAT
    \STATE $\bar{\mathbf{{D}}} = \mathbf{D};$\\
    \STATE $\mathbf{U}_{k} = \left(\displaystyle{\sum}_{j=1}^K \mathbf{H}_{k}\mathbf{D}_{j}\mathbf{D}_{j}^{H}\mathbf{H}_{k}^{H}+\sigma^{2} \mathbf{I}_M \right)^{-1}\mathbf{H}_{k}\mathbf{D}_{k},\; \forall k$;
    \STATE $\mathbf{W}_{k} = \left(\mathbf{I}_L -\mathbf{U}_k^H\mathbf{H}_k\mathbf{D}_k \right)^{-1},\; \forall k$;
    \STATE $\mathbf{A}=\operatorname{blkdiag}\left(\alpha_{1}\mathbf{U}_{1}\mathbf{W}_{1}\mathbf{U}_{1}^{H},\ldots,\alpha_{K}\mathbf{U}_{K}\mathbf{W}_{K}\mathbf{U}_{K}^{H}\right)$;
    \STATE $\mathbf{B}=\operatorname{blkdiag}\left(\alpha_{1}\mathbf{W}_{1}\mathbf{U}_{1}^{H},\ldots,\alpha_{K}\mathbf{W}_{K}\mathbf{U}_{K}^{H}\right)$;	
    \STATE $\mathbf{C} = \sum_{l =1 }^{N}{{\mathbf{d}}}_\ell {\mathbf{h}}_\ell^{H}$ where ${{\mathbf{d}}}_\ell$ is  the $\ell$-th column of $\mathbf{D}^H$;
    \STATE \textbf{for} $n = 1:N$ \textbf{do}\\
    \STATE\quad $a_{n}={\mathbf{h}}_{n}^{H}\mathbf{A} {\mathbf{h}}_{n}$; \\
\STATE\quad 
$\mathbf{b}_{n}= -\mathbf{B}{\mathbf{h}}_{n}+\left(\mathbf{C}-{\mathbf{d}}_n {\mathbf{h}}_{n}^{H} \right)\mathbf{A} {\mathbf{h}}_{n}$;
\STATE {\color{black}{\quad $P^\text{ub}_n = \min\left( P_\text{TX} - \displaystyle{\sum}_{\ell\not =n} \|{\mathbf{d}}_\ell\|_2^2, {\color{black}{P^\text{ub}_\text{PAPC}}}\right)$};}
    \STATE\quad
    ${\mathbf{d}}_{n}=-\mathbf{b}_{n} \cdot {\color{black}{{{\max}}}} \left( \min \left( \frac{1}{a_{n}},\frac{\sqrt{{\color{black}{P^\text{ub}_n}}}}{ \|\mathbf{b}_{n}\|_2} \right), {\color{black}{{{\frac{\sqrt{P^\text{lb}_\text{PAPC}}}{ \|\mathbf{b}_{n}\|_2}}}}} \right)$;
    \STATE\textbf{end for}\\
\STATE $i \leftarrow i + 1$;
    \UNTIL $R(\mathbf{D}) - R(\bar{\mathbf{D}})\leq \epsilon$ or $i > i_\text{max}$.
        \ENSURE $\mathbf{D}= \left[{\mathbf{d}}_{1},\ldots,{\mathbf{d}}_{N} \right]^H$.
    \end{algorithmic}
\end{algorithm}

To complete the description of the algorithm, we need to establish a method for determining a feasible starting precoder. Let us suppose that we are given an initial precoder, denoted as $\changeb{{\mathbf{D}}_\text{init}}$, for example obtained from a conventional precoding method that only satisfies SPC. The per-antenna power of this precoder can be calculated as $\changeb{\bp_\text{init}} = \operatorname{diag}\left( \changeb{{\mathbf{D}}_\text{init} {{{\mathbf{D}}}^H_\text{init}}}\right)$.
\changeb{
If we have another vector of powers $\bp$ that is feasible, then a feasible precoder can be obtained by scaling rows of $\bD_\text{init}$ as:
\begin{equation}
\changeb{ {{\mathbf{D}}} =  \left(\operatorname{Diag}\left({{\bp}\odiv{\bp_\text{init}}}\right) \right)^{1/2}{\bD}_\text{init}.}
\end{equation}
For example, under our assumptions, the vector $\bp=P^\text{lb}_\text{PAPC}\cdot\mathbf{1}_N$ is always feasible. However, empirically we find convergence speed is generally better if we use feasible powers $\bp$ close to $\bp_\text{init}$. 
A number of methods of varying complexity can be used to find a feasible solution reasonably close to $\bp_\text{init}$. 
It is simple to project a power vector $\bp$ to one satisfying PAPC constraints $\bp'$ as:
\begin{equation}
\label{eq_projection}
\bp' = \max \left( \min \left (\changeb{\bp}, P^\text{ub}_\text{PAPC} \right), P^\text{lb}_\text{PAPC} \right) \changeb{.}
\end{equation}
Likewise, projecting a power vector $\bp'$ to a vector $\bp$ satisfying SPC is given by:
\begin{equation}
\label{eq_projection_sum}
\bp=\bp' - \left(\frac{\mathbf{1}_N^T \bp' - P_\text{TX}}{N}\right)^{+} \cdot \mathbf{1}_N .
\end{equation}
Thus, a feasible point close to $\bp_\text{init}$ 
can be found by iteratively applying \eqref{eq_projection} and \eqref{eq_projection_sum} (method of alternating projections) or by using Dykstra's algorithm \cite{combettes2011proximal}.
These methods are guaranteed to converge asymptotically to a feasible solution; to terminate the iterations early at any point with a feasible solution, one can apply  \eqref{eq_projection}, and then if $\bp'$ does not satisfy SPC, enforce SPC via
\begin{equation}
\bp = \frac{P_\text{TX} - N P^\text{lb}_\text{PAPC}}{\mathbf{1}^T_N \bp' - N  P^\text{lb}_\text{PAPC}} \bp' +  \frac{\mathbf{1}^T_N \bp' - P_\text{TX} }{\mathbf{1}^T_N \bp' - N  P^\text{lb}_\text{PAPC}} P^\text{lb}_\text{PAPC}  \cdot \mathbf{1}_N.
\end{equation}
The result also satisfies PAPC as it is a convex combination of $\bp'$ and $P^\text{lb}_\text{PAPC} \mathbf{1}_N$.
}

Finally, it should be mentioned that the computational complexity of the proposed flat WMMSE precoding is similar to the WMMMSE approach in \cite{TomLuo2023}. In particular, for systems with $N \gg K.M \geq K.L$, it can be shown that the computational complexity is dominated by line 8 which has a complexity of $O(N K^2 M L)$ . Therefore, the computational
complexity of each iteration of the proposed flat WMMSE
is linear in the number of BS antennas.

\section{ZF Precoding with Controllable Flatness}
\label{sec:flat_ZF_CF}
In this section, we aim to introduce the concept of controlling the precoding flatness into the design of another widely-used precoding method, ZF. We restrict our attention to the ZF precoder which enforces the ZF structure between all the layers, e.g., even between the layers corresponding to one UE. Here, we assume that the UEs have already designed their \changeb{row-orthonormal} receive combiner $\widetilde{\bU}_k^H \in \mathbb{C}^{L\times M}$ (e.g., by using eigen beamforming), and then the channel vector used for ZF is the effective channel matrix of $\widetilde{\bH}_k = \widetilde{\bU}_k^H \bH_k \in \mathbb{C}^{L \times N}$. Furthermore, we denote the priority weight of each layer $\ell$ as $\widetilde{\alpha}_\ell$, which is set to be equal to the priority weight of the UE that the layer is intended for, i.e., $\widetilde{\alpha}_\ell = \alpha_{\lceil \ell/L \rceil}$.

With these considerations in place, enforcing the ZF structure can mathematically be modeled as:
\begin{equation}\label{ZF_constraint}
\widetilde{\bh}_j^H \widetilde{\bd}_\ell =0, ~~~\forall j  \not =\ell,
\end{equation}
where $\widetilde{\bh}_j^H$ is the $j$-th row of the overall effective channel of $\widetilde{\mathbf{H}} \triangleq \left[\widetilde{\mathbf{H}}_1^T, \ldots, \widetilde{\mathbf{H}}_K^T \right]^T \in \mathbb{C}^{KL \times N}$ and $\widetilde{\bd}_\ell$ is the $\ell$-th column of the digital precoder ${{\mathbf{D}}} \in \mathbb{C}^{N\times KL}$. By employing the ZF structure, all the interference between the layers is omitted and the problem in \eqref{eq:WMMSE_problem} can be rewritten as:
\begin{subequations}
\label{eq:ZF_general}
\begin{eqnarray}
\max_{\mathbf{D}} && \displaystyle{\sum}_{\ell=1}^{KL}  ~\widetilde{\alpha}_\ell \changeb{\log} \left( 1 + \tfrac{1}{\sigma^2} \left|\widetilde{\bh}_\ell^H \widetilde{\bd}_\ell \right|^2  \right) \\ 
\text{s.t.~} &&  \displaystyle{\sum}_{\ell=1}^{KL} \operatorname{Tr}\left( \widetilde{\bd}_\ell \widetilde{\bd}_\ell^H \right)  \leq P_\text{TX}, \\ 
&& \displaystyle{\sum}_{\ell=1}^{KL} \left[\widetilde{\bd}_\ell \widetilde{\bd}_\ell^H \right]_{n,n} \leq P^\text{ub}_\text{PAPC},\forall n=1,\ldots, N,~~~ \\ 
&&  \displaystyle{\sum}_{\ell=1}^{KL} \left[\widetilde{\bd}_\ell \widetilde{\bd}_\ell^H\right]_{n,n} \geq P^\text{lb}_\text{PAPC},\forall n=1,\ldots, N,~~~\\ 
&& \widetilde{\bh}_j^H \widetilde{\bd}_\ell =0, ~~~\forall j  \not =\ell. 
\end{eqnarray}
\end{subequations}
This problem is non-convex. A common practice for transforming such a ZF problem into a convex optimization problem 
is employing SDR which involves matrix lifting together with the rank-one relaxation. To do so, let us define the rank-one covariance matrix of the digital precoder of each stream as $\bQ_\ell = \widetilde{\bd}_\ell \widetilde{\bd}_\ell^H$. By relaxing the rank-one constraint, the problem \eqref{eq:ZF_general} can then be written as follows:
\begin{subequations}
\label{eq:ZF_SDP}
\begin{eqnarray}
\max_{\{\mathbf{Q}_\ell \succeq \mathbf{0}\}_{\forall \ell}} && \displaystyle{\sum}_\ell  ~ \alpha_\ell \changeb{\log} \left( 1 + \tfrac{1}{\sigma^2} \widetilde{\mathbf{h}}^H_{\ell} \mathbf{Q}_\ell \widetilde{\mathbf{h}}_{\ell}\right) \\ 
\text{s.t.~~} &&  \mathbf{1}^T \operatorname{diag} \left( \displaystyle{\sum}_\ell \mathbf{Q}_\ell \right) \leq P_\text{TX},\\ 
&& \displaystyle{\sum}_\ell \left[\mathbf{Q}_\ell\right]_{n,n} \leq P^\text{up}_\text{PAPC},~~~\forall n,\\ 
&& \displaystyle{\sum}_\ell \left[\mathbf{Q}_\ell\right]_{n,n} \geq P^\text{lb}_\text{PAPC},~~~\forall n,~~~~~~~~~~\\ 
&& \widetilde{\mathbf{h}}^H_{j} \mathbf{Q}_\ell \widetilde{\mathbf{h}}_{j}= 0,~~~ \forall j \not= \ell \changeb{.}
\end{eqnarray}
\end{subequations}

The problem in \eqref{eq:ZF_SDP} is a convex optimization problem and we can use off-the-shelf numerical optimization packages to solve it efficiently (e.g., CVX \cite{cvx}). If the optimal solutions for problem \eqref{eq:ZF_SDP}, i.e., $\bQ^\star_\ell, \forall \ell$, are all rank-one, then we can easily recover the design of digital precoders, $\widetilde{\bd}_\ell, \forall \ell$, e.g., by using the eigenvalue decomposition.
In our numerical experiments, we observe that the optimal solutions for problem \eqref{eq:ZF_SDP} are typically rank-one. Similar observations were previously reported by several different works within the context of multi-user precoding, but for slightly different objectives and power constraints \cite{Wisel2008,Ken2017}. Note that even if some $\bQ^\star_\ell$'s are not rank-one, we can use the randomization procedure in \cite{Ken2010} or \textit{Theorem~2} in \cite{Wisel2008} to construct good rank-one solutions based on $\bQ^\star_\ell$'s. However, it is possible that these rank-one solutions may no longer satisfy the power constraints or ZF constraints. In such cases, we can utilize the procedure outlined in Section~\ref{sec:flat_WMMSE_CF} to find an initial feasible precoder that ensures the solution satisfies all three types of power constraints. It is important to note that the ZF structure constraint may still be violated. However, based on our simulations, we have observed that such cases are rare in massive MIMO regime. In the next section, we provide a discussion on the feasibility of the flat ZF precoding in general.

\changeb{While it is possible to employ convex optimization tools and packages to find the global optimal solution for the problem in \eqref{eq:ZF_SDP}, the practical computational complexity of such approaches may still be high. In particular, the computational complexity of solving an SDP problem using an interior-point method, in the worst case, scales cubically with respect to the number of decision variables. For the proposed SDR-based flat ZF method, where the number of optimization variables is $ O(N^2 K L) $, the computational complexity may scale as $ O(N^6 K^3 L^3)$, which can be quite demanding.}

To mitigate the computational complexity in seeking a favorable ZF solution, one can potentially exploit subgradient and/or approximation methods proposed in the literature, e.g., \cite{Rui2010, Pham2018}. However, the computational complexity of these methods still remains relatively high. To address this issue, in the next section, \changeb{we
propose an alternative problem formulation} for flat ZF precoding design that admits a quasi-closed-form solution, enabling efficient solution finding.

\section{Fixed-Relative-Gain Flat ZF Precoding}
\label{sec:flat_ZF_CF_SO}

In this section, our aim is to propose a more computationally efficient algorithm for designing a flat ZF precoder. To achieve this, we first devise an algorithm for designing the flat ZF precoder, assuming that the ratio of the received gains for all the layers is given which we refer to as a fixed-relative-gain (FRG) profile. Subsequently, we discuss how to obtain an appropriate received gain profile that maximizes WSR. Finally, we discuss about the feasibility of finding a flat ZF precoder.     

\subsection{Flat ZF Precoding for a Fixed-Relative-Gain Profile}
\label{sec:flat_ZF_CF_SO_FG}

To simplify the ZF problem in \eqref{eq:ZF_general}, we can specify that the effective received gain for each layer $\ell$ should be proportional to a fixed value $g_\ell \geq 0$. Consequently, we can express the ZF constraint as \changeb{$\widetilde{\bH} \bD = \beta \mathbf{G}$}, where $\bG$ is a diagonal matrix with $[\bG]_{\ell\ell} = g_\ell$. To develop a computationally efficient algorithm, we assume that the received relative gain profile $\bG$ is already given, and our objective is to optimize a flat ZF precoder $\mathbf{D}$ and the common scaling gain factor $\beta$. Assuming that all diagonal elements of $\bG$ are positive\footnote{If there are any zero diagonal elements in $\bG$, we replace them with very small numbers and then perform the inversion.}, we further write $\check{\bH} \bD = \beta \bI_{KL}$, where $\check{\bH} = \bG^{-1}\widetilde{\bH}$. With these considerations, 
 the problem can be re-written as:

\begin{subequations}
\label{eq:ZF_general_so}
\begin{eqnarray}
\max_{\mathbf{D},\beta} && \beta \\
\text{s.t.~} && \check{\bH} \bD = \beta \bI_{KL},\label{subeq:ZF_general_so_gain}\\
&&  \displaystyle{\sum}_{n=1}^{N}
\|{\mathbf{d}}_n\|_2^2 
 \leq P_\text{TX}, \label{subeq:ZF_general_so_tot_ub}\\ 
&& \|{\mathbf{d}}_n\|_2^2  \leq P^\text{ub}_\text{PAPC},\forall n=1,\ldots, N,~~~ \label{subeq:ZF_general_so_ub}\\ 
&& \|{\mathbf{d}}_n\|_2^2  \geq P^\text{lb}_\text{PAPC},\forall n=1,\ldots, N,~~~ \label{subeq:ZF_general_so_lb}
\end{eqnarray}
\end{subequations}
where here the power constraints are written more explicitly in terms of the rows of the precoder $\bD$. 

In this section, we aim to describe how this problem can be efficiently solved, first without LB constraints \eqref{subeq:ZF_general_so_lb}, and then with these bounds included.
In the proposed algorithm, we will use the following Lagrangian function:

\begin{eqnarray} 
{\cal L}\left( \beta, \bD; \bZ,\boldsymbol{\lambda},\boldsymbol{\theta} ,\mu\right)  =  \beta + {2}\displaystyle{\sum_{j,\ell}} \changeb{ \mbox{Re}\left( [\bZ]_{j,\ell}^*\left( \check{\bh}_j^H {\widetilde{\bd}}_\ell -\beta [\bI]_{j,\ell}\right)\right) }\nonumber \\
 + \mu \left( P_\text{TX} - \sum_{n=1}^N \| {\mathbf{d}}_n \|_2^2  \right) \hspace{47pt} \nonumber \\
+ \sum_{n=1}^N \lambda_n \left(P^\text{ub}_\text{PAPC} -  \| {\mathbf{d}}_n\|_2^2 \right) \hspace{41pt} \nonumber \\
+ \sum_{n=1}^N \theta_n \left( \|{\mathbf{d}}_n\|_2^2 - P^\text{lb}_\text{PAPC} \right), \hspace{41pt} \nonumber
\end{eqnarray}
where $\check{\bh}_j^H$ is the $j$-th row of $\check{\bH}$, and $\bZ,\boldsymbol{\lambda},\boldsymbol{\theta} ,\mu$ are Lagrangian multipliers (i.e., dual variables) corresponding to different constraints in problem \eqref{eq:ZF_general_so}.
The dual variable $\bZ$ is a complex matrix, while the other multipliers $\mu, \boldsymbol{\lambda}$, and $\boldsymbol{\theta}$, associated with inequality constraints, are real and non-negative.

\subsubsection{Without LB-PAPCs}
In the absence of the LB constraints \eqref{subeq:ZF_general_so_lb}, the problem becomes convex as the objective is linear and all the constraints are convex. Hence, the primal-dual approach offers a convenient iterative method to achieve a global maximum. In this scenario, the Lagrangian is modified by eliminating the terms involving multipliers $\theta_n$.

The dual function ${\cal D}\left( \bZ,\boldsymbol{\lambda}, \mu\right)$ is defined as the maximum of the Lagrangian function over the primal variables $\beta$ and $\bD$, for fixed dual variables.
Taking derivatives with respect to $\beta$, we see that the dual function is infinite unless \changeb{$\tr\left(\mbox{Re}\left({\bZ}\right)\right)=1/2$}.
As the Lagrangian function is a concave quadratic in $\bD$, the dual function is obtained where the derivative with respect to $\bD$ is equal to zero. The derivative is expressed in matrix form as $2\check{\bH}^H \bZ - 2\left(\mu \bI_{N} + \boldsymbol{\Lambda}\right) \bD$ where $\boldsymbol{\Lambda}$ is a diagonal matrix with $[\boldsymbol{\Lambda}]_{nn} = \lambda_n$, and thus the optimizing precoder is:
\begin{equation}
\bD^\star = \left(\mu \bI_N + \boldsymbol{\Lambda} \right)^{-1}\check{\bH}^H \bZ.
\end{equation}
In summary, the dual function has the following closed form:
If \changeb{$\tr\left(\mbox{Re}\left({\bZ}\right)\right)\neq 1/2$}, we have ${\cal D}(\bZ,\boldsymbol{\lambda},\mu) = \infty$; otherwise, we have:
\begin{eqnarray}
{\cal D}(\bZ,\boldsymbol{\lambda},\mu) = {\cal L}\left( \beta,\bD^\star; \bZ,\boldsymbol{\lambda},\boldsymbol{0},\mu \right).
\end{eqnarray}
As is well known for convex problems, the primal maximization problem can be solved by minimizing the dual function over the dual variables \cite{boyd2004convex}. This approach is especially convenient in our case because the dual variables are only subject to non-negativity constraints, and also because the dual function can be minimized component-wise in quasi closed form.

Consider minimization of the dual function over $\bZ$ for fixed values of $\mu$ and $\boldsymbol{\lambda}$. The derivative of the dual function with respect to $\bZ$ is given by the associated constraint violation $\changeb{2} \check{\bH} \bD^\star - \changeb{2}\beta \bI_{KL}$. The critical point (zero derivative) is thus obtained in closed form with:
\begin{equation}
\bZ = \beta^\star \left( \check{\bH} \left(\mu \bI_N + \boldsymbol{\Lambda}\right)^{-1} \check{\bH}^H \right)^{-1},
\label{eq:opt_U}
\end{equation}
where $\beta^*$ must be chosen to get \changeb{$\tr\left({\bZ}\right)=1/2$ (noting that as $\bZ$ in (\ref{eq:opt_U}) is non-negative definite, its trace is real).}

For fixed $\bZ$ and $\mu$, the derivative of the dual function with respect to $\lambda_n$ is the violation $P^\text{ub}_\text{PAPC}-\|\bd^\star_n\|_2^2$. If we define $\bV\triangleq\check{\bH}^H \bZ$ with $\bv_n$ being its $n$-th row, then we can write $\|\bd^\star_n\|_2^2= (\mu + \lambda_n)^{-2} \|\bv_n\|_2^2$. It can be shown that the minimization of the dual function with respect to $\lambda_n \geq 0$ is achieved when:
\begin{equation}\label{eq:opt_lambda}
\changeb{\lambda_n = \left( \frac{\changebb{\|\bv_n\|_2}}{\sqrt{P^\text{ub}_\text{PAPC}}} - \mu \right)^+.}
\end{equation}
Finally, the derivative of the dual function with respect to $\mu$ is given by $P_\text{TX}-\sum_n \| \bd^\star_n \|_2^2$. Plugging the optimized $\lambda_n$ for each value of $\mu$, we have to get $\|\bd_n\|_2^2= \min \left( \mu ^{-2} \|\bv_n\|_2^2, P^\text{ub}_\text{PAPC} \right)$. \changeb{Assuming $N P^\text{ub}_\text{PAPC} > P_\text{TX}$,} the dual function can be minimized with respect to $\mu$ by one dimensional search to solve the equation:
\begin{equation}
\sum_n \min \left( \mu ^{-2} \|\bv_n\|_2^2, P^\text{ub}_\text{PAPC} \right) = P_\text{TX}.
\label{eq:opt_mu}
\end{equation}
\changeb{Otherwise, we can take $\mu=0$ and the total power constraint is redundant and can be ignored.}

By iteratively optimizing with respect to $\bZ$ using the solution in \eqref{eq:opt_U}, \changeb{with respect to $\lambda_n$ using \eqref{eq:opt_lambda}, and with respect to $\mu$ using \eqref{eq:opt_mu},} the dual function can be minimized iteratively towards the global minimum. In each step that we optimize with respect to $\bZ$, the resulting precoder $\bD^\star$ satisfies the ZF condition $\check{\bH}\bD^\star = \beta^\star \bI_{KL}$. However, it may not satisfy the power constraints. To address this issue, we can obtain a primal feasible solution by scaling down $\bD^\star$ and $\beta^\star$ by a common factor $\nu \leq 1$ such that $\bD^\star$ meets the power constraints. In each iteration we obtain a feasible solution $\nu \bD^\star$ with the common receive gain factor $\nu \beta^\star$, and we know that the optimal common gain factor is no more than $\beta^\star$. The iteration can be terminated when $\nu$ is sufficiently close to 1.

\subsubsection{With LB-PAPCs}

When LB-PAPCs are present, the constraints are no longer convex, and theoretical statements are harder to make. However, the dual function still provides upper bounds and typically leads to good primal solutions as well. Extending the dual minimization method to this problem gives a tractable approach that seems to work well in practice, as demonstrated by numerical results later.

In this case, the dual function can again be obtained in closed form. Note that, the Lagrangian function is now concave only if $\mu + \lambda_n - \theta_n \geq 0$. The optimizing precoder is given by:
\begin{equation}
\bD^\star = \left(\mu \bI_N + \boldsymbol{\Lambda} - \boldsymbol{\Theta}\right)^{-1}\check{\bH}^H \bZ,
\end{equation}
where $\boldsymbol{\Theta}$ is a diagonal matrix with $[\boldsymbol{\Theta}]_{nn} = \theta_n$. If \changeb{$\tr\left(\mbox{Re}\left({\bZ}\right)\right)\neq 1/2$,} or if $\mu + \lambda_n - \theta_n < 0$ for any $n$, we have ${\cal D}(\bZ,\boldsymbol{\lambda},\boldsymbol{\theta},\mu) = \infty$. Otherwise, we have:
\begin{eqnarray}
{\cal D}(\bZ,\boldsymbol{\lambda},\boldsymbol{\theta},\mu) = {\cal L}\left( \beta,\bD^\star; \bZ,\boldsymbol{\lambda},\boldsymbol{\theta},\mu \right).
\end{eqnarray}

\changeb{Similar to the case without LB-PAPCs, when LB-PAPC's are present the dual can again be minimized separately for $\bZ$ and for $\mu$; the variables $\lambda_n$ can also be optimized separately from $\bZ$ and $\mu$, but jointly with $\theta_n$.} The optimization over $\bZ$ yields:
\begin{equation}
\bZ = \frac{\left( \check{\bH} \left(\mu \bI + \boldsymbol{\Lambda}-\boldsymbol{\Theta}\right)^{-1} \check{\bH}^H \right)^{-1}}{\changeb{2}\tr \left( \left( \check{\bH} \left(\mu \bI + \boldsymbol{\Lambda}-\boldsymbol{\Theta}\right)^{-1} \check{\bH}^H \right)^{-1} \right)}. 
\end{equation}
\changeb{In the joint optimization of $\lambda_n$ and $\theta_n$, \changebb{it is} straightforward to show that if $\changebb{\|\bv_n\|_2} \geq \mu \sqrt{P^\text{ub}_\text{PAPC}}$, we have $\lambda_n\geq0$ and $\theta_n=0$, if $\changebb{\|\bv_n\|_2} \leq \mu \sqrt{P^\text{lb}_\text{PAPC}}$, we have $\lambda_n = 0$ and $\theta_n \geq 0$, and that otherwise $\lambda_n = \theta_n = 0$.
As a result, the jointly optimized variables satisfy \eqref{eq:opt_lambda} and
\begin{equation}
    \changeb{\theta_n =  \left( \mu  - \frac{\changebb{\|\bv_n\|_2}}{\sqrt{P^\text{lb}_\text{PAPC}}} \right)^+.}
\end{equation}}

Finally, the optimization over $\mu$ can be done by one dimensional search for a solution to the following equation:
\begin{equation}
\sum_n \max\left( \min \left( \mu ^{-2} \|\bv_n\|_2^2, P^\text{ub}_\text{PAPC} \right) , P^\text{lb}_\text{PAPC} \right) = P_\text{TX}.
\label{eq:search_for_mu}
\end{equation}
\changeb{As with (\ref{eq:opt_mu}), we take $\mu=0$ if $NP^\text{ub}_\text{PAPC} ~\changebb{\leq}~ P_\text{TX}$. Also, if
$NP^\text{lb}_\text{PAPC} >  P_\text{TX}$ the constraints are infeasible and there is no solution to (\ref{eq:search_for_mu}).}

Obtaining a feasible primal solution from $\bD^\star$ becomes more challenging when LB-PAPCs are present. This is because scaling down $\bD^\star$ to satisfy upper bound constraints (i.e., SPC and UB-PAPCs) may result in the violation of some LB-PAPCs. A useful heuristic to address this problem is to carry out the dual function computations with a more stringent LB constraint $P^\text{lb}_\text{PAPC}+\epsilon_\text{lb}$. This provides some leeway for a dual optimizer $\bD^\star$ to be scaled down to $\nu D^\star$, to obtain a solution that still respects the true LB constraint $P^\text{lb}_\text{PAPC}$.

The summary of the FRG-flat ZF algorithm is illustrated in Algorithm~\ref{alg_fixed_gain_ZF}. Note that when LB-PAPCs are present, there is no guarantee that a feasible solution exists, or that the primal dual approach finds a feasible solution. However, it is observed that the primal-dual method performs well with typical wireless channels, especially in massive MIMO regime. Further discussions on feasibility are provided in Section~\ref{sec:flat_ZF_feasibility}. 
\changeb{Finally, it should be mentioned that for given $\widetilde{\bH}$ and $\bG$, the computational complexity of the proposed FRG-Flat ZF is given by $O(N  K^2 L^2)$ , which is linear in terms of the number of BS antennas. This makes it much more efficient compared to the SDR-based flat ZF presented in the previous section.}

\begin{algorithm}[!tb]
    \renewcommand{\algorithmicrequire}{\textbf{Input:}} 
    \renewcommand{\algorithmicensure}{\textbf{Output:}}
    \caption{FRG-Flat ZF Precoding}  \label{alg_fixed_gain_ZF}
    \begin{algorithmic}[1]
    \REQUIRE System dimensions $N$, $K$, $M$, and $L$; effective channel matrix $\widetilde{\mathbf{H}}$; relative gain profile $\bG$; power constraints $P_\text{TX}$, $P^\text{ub}_\text{PAPC}$, and $P^\text{lb}_\text{PAPC}$; LB tightening parameter $\epsilon_\text{lb}$; convergence threshold $\epsilon$; maximum number of iterations $i_\text{max}$.
    \STATE Set $P^\text{lb}_\text{PAPC} = P^\text{lb}_\text{PAPC} + \epsilon_\text{lb}$ and $\check{\bH} = \bG^{-1}\widetilde{\bH}$.
    \STATE Initialize $\bZ = \tfrac{\left( \check{\bH} \check{\bH}^H \right)^{-1}}{\changeb{2}\tr \left( \left( \check{\bH} \check{\bH}^H \right)^{-1} \right)}$ and $i=1$.
    \REPEAT
    \STATE Optimize $\mu$ by solving the equation \eqref{eq:search_for_mu}, e.g., by performing a one-dimensional search;
    \STATE \changeb{$\lambda_n = \left( \tfrac{\changebb{\|\bv_n\|_2}}{\sqrt{P^\text{ub}_\text{PAPC}}} - \mu \right)^+$};
    \STATE \changeb{$\theta_n = \left( \mu  - \tfrac{\changebb{\|\bv_n\|_2}}{\sqrt{P^\text{lb}_\text{PAPC}}} \right)^+$};
    \STATE $\bZ = \tfrac{\left( \check{\bH} \left(\mu \bI + \boldsymbol{\Lambda}-\boldsymbol{\Theta}\right)^{-1} \check{\bH}^H \right)^{-1}}{\changeb{2}\tr \left( \left( \check{\bH} \left(\mu \bI + \boldsymbol{\Lambda}-\boldsymbol{\Theta}\right)^{-1} \check{\bH}^H \right)^{-1} \right)}$;
    \STATE $\bD^\star = \left(\mu \bI_N + \boldsymbol{\Lambda} - \boldsymbol{\Theta} \right)^{-1}\check{\bH}^H \bZ$;
    \STATE Find the maximum $\nu \leq 1$ such that $\nu \bD^\star$ satisfies SPC and UB-PAPCs.
    \STATE $i \leftarrow i + 1$;
    \UNTIL $|\nu - 1| \leq \epsilon$ or $i > i_\text{max}$.
        \ENSURE $\nu \mathbf{D}^\star$.
    \end{algorithmic}
\end{algorithm}

\subsection{Received Gain Profile Optimization}
\label{sec:flat_ZF_CF_SO_AGP}

In the previous subsection, we discussed how to maximize a common gain $\beta$ by designing a digital precoder that satisfies the ZF and power constraints, given a relative gain profile $\bG$. In this section, we explain the process of selecting an appropriate relative gain profile. Specifically, we propose obtaining the optimal receiver gains for maximizing WSR under only SPC, and then using this as the relative gain profile for optimizing the precoder based on the proposed algorithm in Algorithm~\ref{alg_fixed_gain_ZF}.

As shown in \cite{Cornelis2023WSRZF}, the ZF solution to the WSR maximization problem under SPC is given by the channel inversion followed with a power allocation as:
\begin{equation}
    \bD_\text{SPC} = \widetilde{\bH}^H \left( \widetilde{\bH} \widetilde{\bH}^H\right)^{-1} \sqrt{\boldsymbol{\Gamma}},
    \label{eq:ZF_SPC_WF}
\end{equation}
where $\boldsymbol{\Gamma}$ is the diagonal power allocation matrix with $[\boldsymbol{\Gamma}]_{\ell \ell} = \gamma_\ell$. Further, letting $\tau_\ell = \left[\left( \widetilde{\bH} \widetilde{\bH}^H\right)^{-1}\right]_{\ell,\ell}$, the optimal power allocations is given by the following water-filling solution \cite{Cornelis2023WSRZF}:
\begin{equation}
    \gamma_\ell = \sigma^2 \left( \psi \frac{\widetilde{\alpha}_\ell}{\tau_\ell} -1 \right)^{+}
\end{equation}
where the water level $\psi$ is determined from:
\begin{equation}
    \changeb{\sum_\ell \sigma^2 \left( \psi {\widetilde{\alpha}_\ell} - {\tau_\ell} \right)^{+} = P_\text{TX}.}
\end{equation}

For the ZF precoder under SPC in \eqref{eq:ZF_SPC_WF}, the received signal gain for layer $\ell$ is $\sqrt{\gamma_\ell}$. In this paper, we utilize these received gains, obtained by considering only SPC, to construct the target received gain profile for our problem that incorporates all three types of power constraints:
\begin{equation}
    g_\ell = \frac{\sqrt{\gamma_\ell}}{\sum_j \sqrt{\gamma_j}}.
\end{equation}

\subsection{Discussions on Feasibility of Flat ZF Precoding}
\label{sec:flat_ZF_feasibility}

It is important to note that not all channel matrices admit a ZF precoder with arbitrarily close UB- and LB-PAPCs. For example, in the case of a square effective channel matrix with $KL = N$, the channel matrix has a unique inverse. This means that if the relative gain profile is fixed, there is only one feasible solution to the equation $\check{\bH}\bD= \beta \bI_{KL}$ for each value of $\beta$. As a result, the relative ratios of the various row norms $\changebb{\| \mathbf{d}_n \|_2}$ which gives the per-antenna powers are fixed and typically not all equal. This implies that the power allocation across the different transmit antennas is not uniform and depends on the specific relative gains profile. In general, the larger $N$ is compared with $KL$, the more degrees of freedom there are to achieve similar transmit powers on all antennas. In our experience with realistic channel matrices with $N \gg KL$, arbitrarily flat profiles can typically be achieved.

However, examples can be constructed with $N \gg KL$ for which the antenna powers cannot be flat. As an example, consider a $KL\times N$ matrix $\check{\bH}$ such that, for all $\ell$,  $|[\check{\bH}]_{\ell,1}|\geq \delta > 0$ and such that $|[\check{\bH}]_{\ell,n}| \leq \xi$ for $n>1$ . Take $\beta = 1 $ without loss of generality, and let $\bD$ satisfy the ZF condition $\check{\bH}\bD=\bI_{KL}$.  Let $\zeta^2 = \max_n \| \mathbf{d}_n \|_2^2$, which implies that $|[\bD]_{n,k}|<\zeta$ for all $n,k$. 
From the zero-forcing condition, for any $j\neq k$ we have
\begin{equation}
[\check{\bH}]_{j,1}[\bD]_{1,k} = - \sum_{n>1} [\check{\bH}]_{j,n} [\bD]_{n,k},
\end{equation}
which implies
$|[\bD]_{1,k}| \leq (N-1) \xi \zeta \delta^{-1}$.
Then the ratio of minimum antenna power to maximum antenna power, for any zero forcing precoder, is upper bounded by
\begin{equation}
\frac{ \min_n  \| \mathbf{d}_n \|_2^2}{ \max_n  \| \mathbf{d}_n \|_2^2} \leq  
\frac{ \| \mathbf{d}_1 \|_2^2}{\zeta^2} \leq
KL(N-1)^2 \xi^2 \delta^{-2} := \rho
\end{equation}
Thus a ZF precoder for $\check{\bH}$ can satisfy per-antenna power constraints only if $\rho 
 \geq P^\text{lb}_\text{PAPC} / P^\text{ub}_\text{PAPC}$. 
For any channel dimensions $N$, $L$, and $K$, and any finite ratio $P^\text{lb}_\text{PAPC} / P^\text{ub}_\text{PAPC} > 0$, we can construct an infeasible problem by choosing $\check{\bH}$ with sufficiently small $\xi/\delta$. 

\section{Simulations}
\label{sec:numerical}

In this section, we assess the performance of various proposed flat precoding methods and compare it with the performance of \changeb{conventional ZF \cite{Cornelis2023WSRZF} and WMMSE \cite{shi2011iteratively} methods}. Before presenting the simulation settings and numerical results, we briefly describe how we calculate power consumption and outline the required PA properties \changeb{as well as} the considered power constraints for each method.

\subsection{Discussion on Power Consumption}
\label{subsec:power_consumption}
To characterize power consumption in the downlink, we adopt a simplified version of the model in \cite{Stefan_2023_Mag} as follows: 
\begin{equation} 
P = P_\text{PA} + N  P_\text{AE}, 
\end{equation} 
where $P_\text{PA}$ represents the overall power consumed by the PAs, and $P_\text{AE}$ denotes the remaining power consumption per antenna element (AE) which is mainly determined by the power consumption of the driver, digital frontend (DFE), and layer-1 processing. The overall PA power consumption is essentially the summation of powers of the PAs, i.e.,
\begin{equation}
P_\text{PA} = \epsilon_\text{IL}  \sum_{n=1}^N \frac{P^\text{TX}_{n}}{\eta_{n}},
\end{equation}
where $\epsilon_\text{IL}$ is the antenna network insertion loss, ${P_n^\text{TX}}$ is the transmitted power from antenna $n$, and $\eta_{n}$ is its corresponding PA power efficiency. We adopt a two-stage Doherty GaN PA model in which the efficiency of the $n$-th PA is \cite{Doherty2016}:
\begin{equation}
\label{eq:Doherty_model}
\changeb{\eta_n = \eta_\text{max} 10^{-(P_\text{sat}^\text{dB} - P^\text{Out,dB}_n -6)^+/20},}
\end{equation}
where $\eta_\text{max}$ is the maximum PA power efficiency, $P_\text{sat}^\text{dB}$ is the PA saturation power, and $P^\text{Out,dB}_n = P^\text{TX,dB}_{n} + \epsilon_\text{IL}^\text{dB}$ is the PA output power. In this model, it is apparent that the \changeb{PA efficiency drops for antenna elements with low output powers.} 

\changeb{Furthermore, the maximum possible saturation power $P_{\text{sat}}^{\text{max,dB}}$ for GaN PAs at $f_c$ GHz is determined by $38 - 16 \log_{10}(f_c)$ dBW \cite{Wang_2021_PA_survey}. Subsequently, the required saturation power $P_\text{sat}^\text{dB} ~ \changebb{\leq}~  P_\text{sat}^\text{max,dB}$ for a beamforming method is calculated based on the maximum PA output power needed for that method, plus an additional back-off power of $P_\text{backoff}^\text{dB}$ to accommodate potential variations in signal/waveform power.}

Now, we discuss the power constraints considered for the baselines as well as the proposed flat precoding methods.
In practice, the maximum permissible radiated power $P_\text{TX}$ is determined based on the existing regulations (e.g., FCC regulations on EIRP) and hardware limitations. In addition, in order to operate in the linear regime of the PA, there is an actual limitation on the maximum feasible radiated power of each antenna which can be calculated as:
\begin{equation}
    P^\text{ub (actual)}_\text{PAPC} = 10^{\left(P_\text{sat}^\text{max,dB} - P_\text{backoff}^\text{dB} - \epsilon_\text{IL}^\text{dB} \right)/10}.
\end{equation}
For the conventional methods, we impose a SPC based on $P_\text{TX}$ and a UB-PAPC based on $P^\text{ub (actual)}_\text{PAPC}$. Furthermore, to demonstrate that the proposed flat beamforming methods can indeed control the level of power flatness across antennas, we use the following PAPC bounds for these methods:
\begin{subequations}
    \begin{eqnarray}
    P^\text{ub}_\text{PAPC} &=& \Delta_\text{p} \changeb{\cdot} \frac{P_{TX}}{N},\\
    P^\text{lb}_\text{PAPC} &=& \frac{1}{\Delta_\text{p}} \changeb{\cdot} \frac{P_{TX}}{N},
    \end{eqnarray}
\end{subequations}
in which the radiated powers of each antenna is enforced to fluctuate within an interval of size $2\Delta_\text{p}^\text{dB}$
around the average radiated power per antenna (i.e., $\tfrac{P_\text{TX}}{N}$).

\begin{figure*}[t]
  \centering
  \begin{subfigure}{2.8in}
    \includegraphics[width=\textwidth]{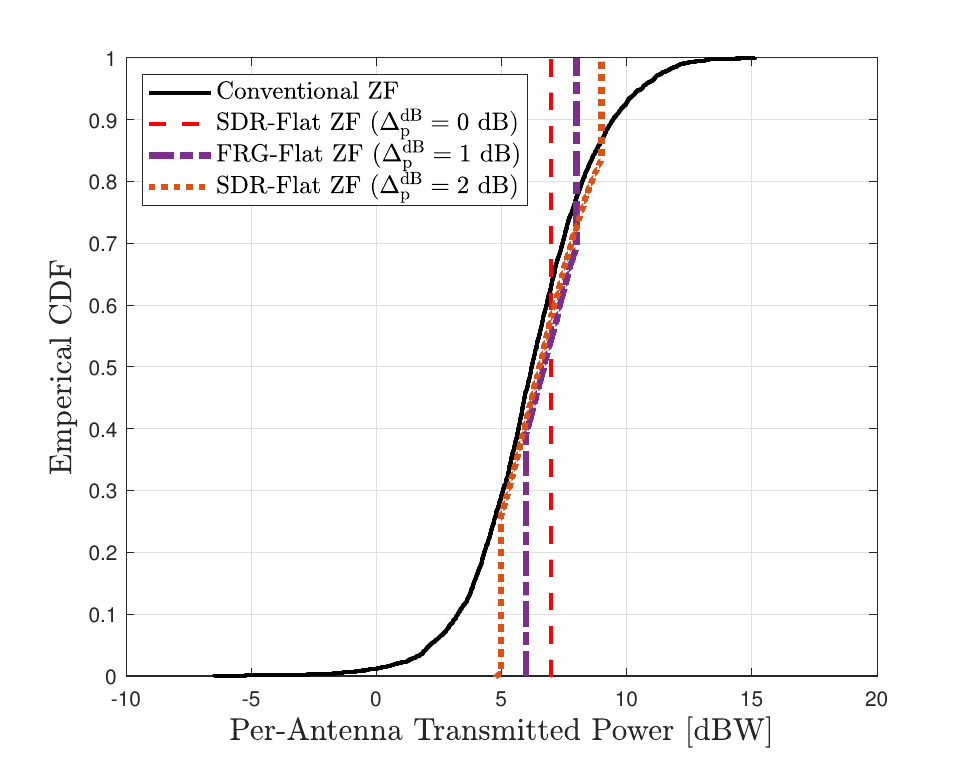}
    \caption{ZF precoding schemes.}
  \end{subfigure}%
  \hfil
  \begin{subfigure}{2.8in}
    \includegraphics[width=\textwidth]{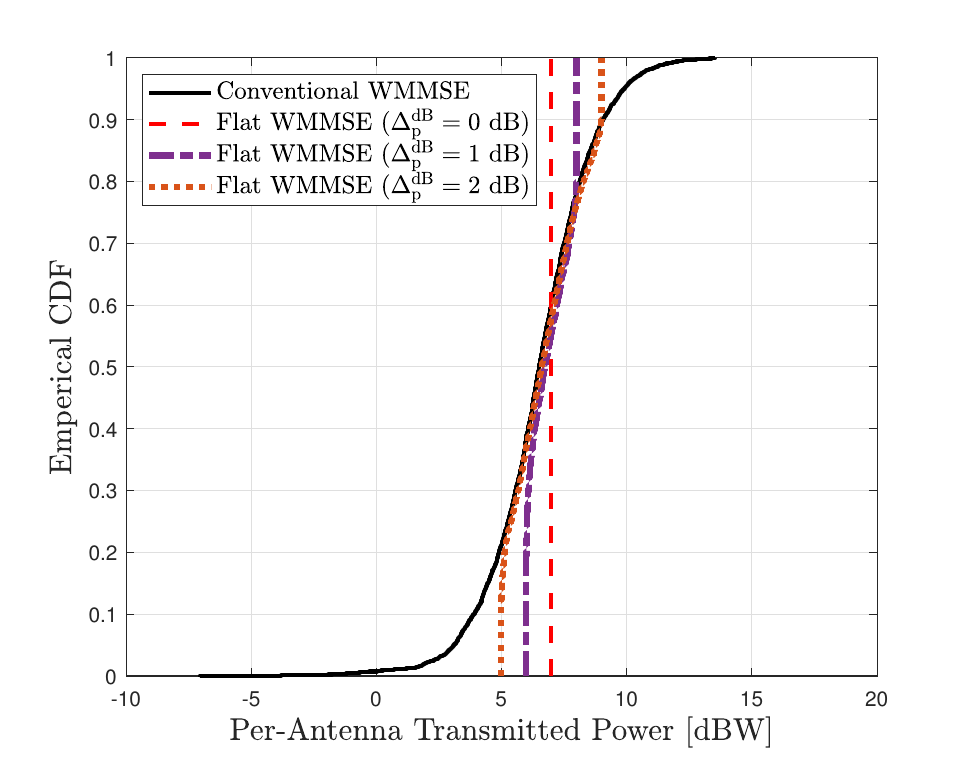}
    \caption{WMMSE precoding schemes.}
  \end{subfigure}
  \caption{The CDF curve of per-antenna transmitted power.}
  \label{fig:exp1}
\end{figure*}

\begin{table*}[ht]
  \centering
  \begin{tabular}{|p{3cm}|p{2cm}|p{2.4cm}|p{2.4cm}|p{2.2cm}|p{2.2cm}|}
    \hline
    \centering{Precoder\\Type} & \centering{Avg. Sum Rate [Gbit/sec]} & \centering{PA Saturation Power [dBW]} & \centering{Power Consumption [W]} & \centering{Energy Efficiency [Mbit/Joule]} & {\centering{Avg. CPU Time\\
    ~~~~~~~ [sec]}}
    \\
    \hline
    \centering FRG-Flat ZF ($\Delta_\text{p}^\text{dB} = 0$) & \centering 18.27 & \centering 15.99 & \centering 569 & \centering 25.91 & {\centering ~~~~~~~~~0.005}
    \\
    \hline
    \centering FRG-Flat ZF ($\Delta_\text{p}^\text{dB} = \changeb{1}$) & \centering 18.50 & \centering 16.99 & \centering 635 & \centering 23.99 & {\centering ~~~~~~~~~0.005}
    \\
    \hline
    \centering FRG-Flat ZF ($\Delta_\text{p}^\text{dB} = \changeb{2}$) & \centering 18.64 & \centering 17.99 & \centering 705 & \centering 22.16 & {\centering ~~~~~~~~~0.005}
    \\
    \hline
    \centering SDR-Flat ZF ($\Delta_\text{p}^\text{dB} = 0$) & \centering 18.38 & \centering 15.99 & \centering 569 & \centering 26.07 & {\centering ~~~~~~~~~71.4}
    \\
    \hline
    \centering SDR-Flat ZF ($\Delta_\text{p}^\text{dB} = \changeb{1}$) & \centering 18.59 & \centering 16.99 & \centering 635 & \centering 24.11 & {\centering ~~~~~~~~~73.3}
    \\
    \hline
    \centering SDR-Flat ZF ($\Delta_\text{p}^\text{dB} = \changeb{2}$) & \centering 18.71 & \centering 17.99 & \centering 705 & \centering 22.24 & {\centering ~~~~~~~~~72.2}
    \\
    \hline
    \centering Conventional ZF & \centering 18.80 & \centering 24.14 & \centering 1394 & \centering 12.29 & {\centering ~~~~~~~~0.0009}
    \\
    \hline
    \centering Flat WMMSE ($\Delta_\text{p}^\text{dB} = 0$) & \centering 19.03 & \centering 15.99 & \centering 569 & \centering 27.00 & {\centering ~~~~~~~~~0.22}
    \\
    \hline
    \centering Flat WMMSE ($\Delta_\text{p}^\text{dB} = 1$) & \centering 19.90 & \centering 16.99 & \centering 634 & \centering 25.81 & {\centering ~~~~~~~~~0.22}
    \\
    \hline
    \centering Flat WMMSE ($\Delta_\text{p}^\text{dB} = 2$) & \centering 20.25 & \centering 17.99 & \centering 705 & \centering 24.05 & {\centering ~~~~~~~~~0.22}
    \\
    \hline
    \centering Conventional WMMSE & \centering 20.39 & \centering 22.52 & \centering 1173 & \centering 15.58 & {\centering ~~~~~~~~~0.10}
    \\
    \hline
  \end{tabular}
  \caption{Performance-complexity comparison for different precoding methods.}
  \label{tab:exp1}
\end{table*}

\subsection{Numerical Results}

Throughout the simulations, we use the 3GPP urban micro (UMi) 3D scenarios described in \cite{3GPP_channel_model}. Further, we examine communication at 7GHz over a 400MHz bandwidth, focusing on a single sector within a hexagonal cellular system with a BS-to-BS distance of 200m. To simplify the analysis, we disregard inter-cell interference. The BS is assumed to be equipped with a planar array of dimensions $8\times N_\text{col}$, with a row spacing of $0.7\lambda$ and column spacing of $0.5\lambda$, where $\lambda$ denotes the wavelength. Additionally, each UE is equipped with a horizontal uniform linear array with half-wavelength antenna spacing.
In the simulations, we use the following parameters: $P_\text{AE} = 4.25$Watt, $\epsilon_\text{IL}^\text{dB} = 2$dB, $P_\text{backoff}^\text{dB} = 7$dB, $P_{\text{sat}}^{\text{max,dB}} = 38 - 16 \log_{10}(7) = 24.47$dBW, $P_\text{TX} = 160$Watt, and $\eta_\text{max} = 0.5$. Further, the power spectral density of the noise is set to $-165$dBm/Hz.

\begin{figure}[t]
  \centering
  \begin{subfigure}{2.8in}
    \includegraphics[width=\textwidth]{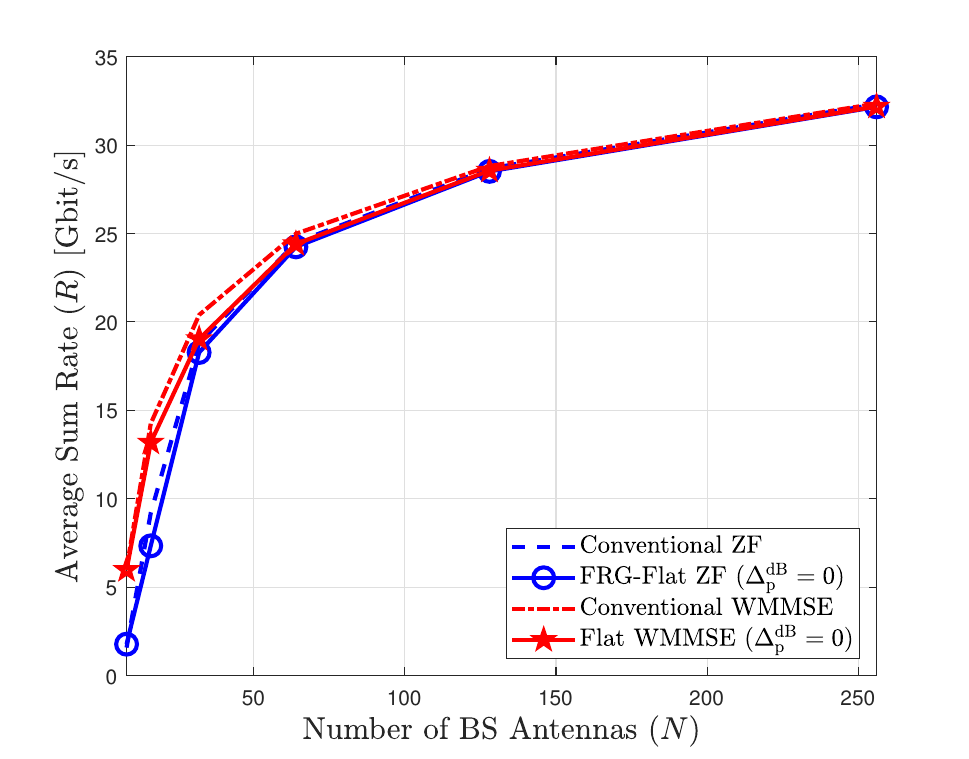}
    \caption{Average sum rate versus $N$.}
    \label{subfig:exp2_1}
  \end{subfigure}\\%
  \begin{subfigure}{2.8in}
    \includegraphics[width=\textwidth]{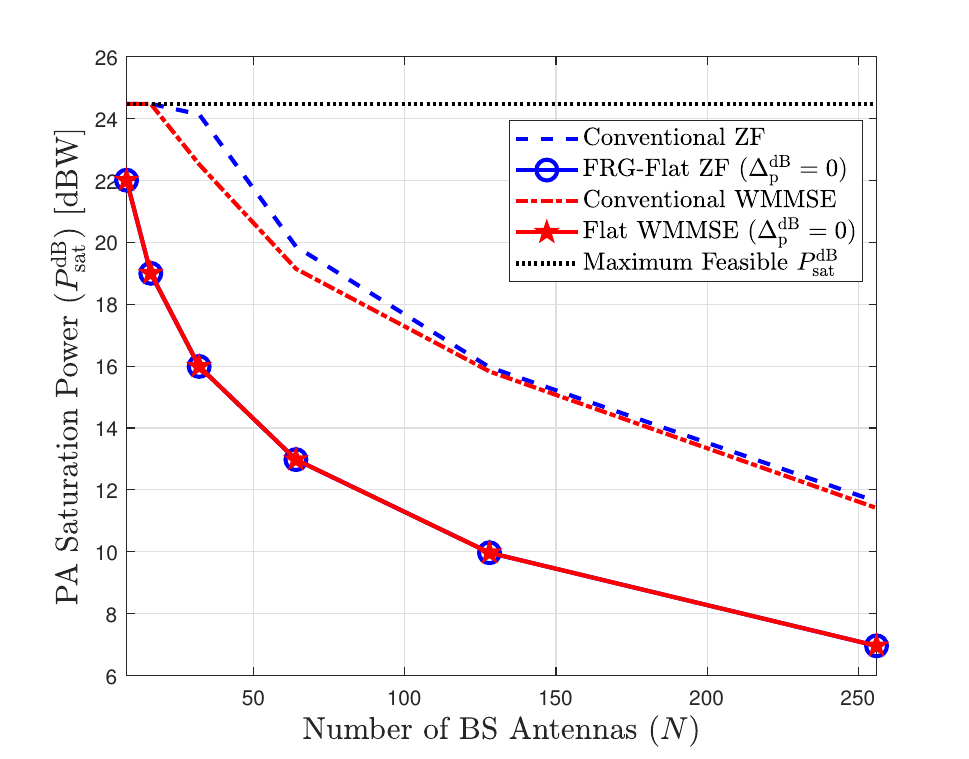}
    \caption{PA saturation power versus $N$.}
    \label{subfig:exp2_2}
  \end{subfigure}\\
  \begin{subfigure}{2.8in}
    \includegraphics[width=\textwidth]{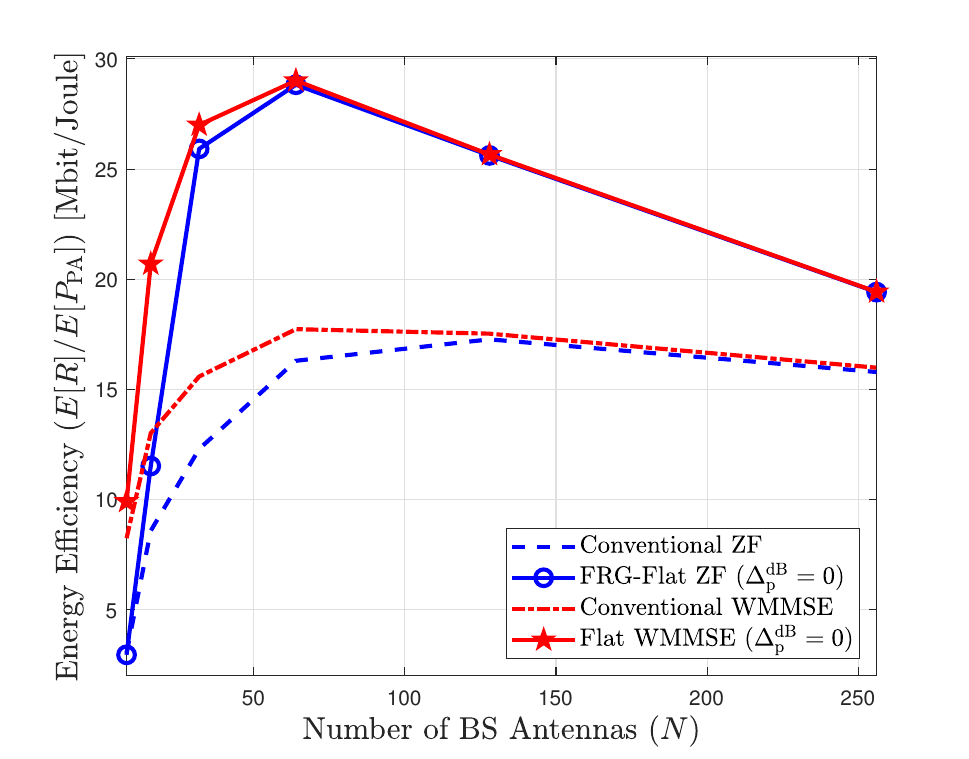}
    \caption{Energy efficiency versus $N$.}
    \label{subfig:exp2_3}
  \end{subfigure}
  \caption{The performance comparison between different methods in a MU-MISO system with $K=8$ single-antenna UEs and $N$ BS antennas.}
  \label{fig:exp2}
\end{figure}

In the first experiment, we consider a scenario in which a BS with $N=32$ antennas serves $K=8$ single-antenna UEs with equal priority weights, i.e., $M=1$ and $\alpha_k =1, \forall k$. For the proposed flat BF methods that have controllable flatness feature, we examine some choices of flatness by setting $\Delta_\text{p}^\text{dB}$ from \changeb{$\{0, 1, 2\}$}. Fig.~\ref{fig:exp1} plots the the empirical CDF of the per-antenna transmitted power for different WMMSE and ZF methods. In particular, it shows that the proposed flat WMMSE, as well as the two proposed flat ZF methods, can effectively maintain the level of per-antenna radiated power fluctuations within a range of $\Delta_\text{p}^\text{dB}$ from the per-antenna average radiated power.

Furthermore, for this simulation setting, Table~\ref{tab:exp1} illustrates the performance of different methods in terms of average sum rate, PA power consumption, energy efficiency, required PA saturation power, and average CPU time. The first observation from Table~\ref{tab:exp1} is that the proposed flat precoding methods can in general approach the performance of their corresponding conventional baseline, while significantly reduce the power consumption and the PA saturation level (hence the PA size). It can also be seen that for flat precoding methods, by increasing $\Delta_\text{p}^\text{dB}$, a higher sum rate can be achieved; however, the gain may be marginal as in this simulation. This observation, along with the fact that the power consumption as well as the PA saturation power reduce with smaller choice of $\Delta_\text{p}^\text{dB}$, leads us to conclude that a completely flat precoder with $\Delta_\text{p}^\text{dB} = 0$ is preferred for the considered PA technology. Further, the proposed FRG-flat ZF method presented in Section~\ref{sec:flat_ZF_CF_SO} can design a flat precoder with a comparable performance to the SDR-Flat ZF presented in Section~\ref{sec:flat_ZF_CF}, while it can significantly reduce the computational complexity. Therefore, for the rest of the simulations related to ZF structure, we mainly focus on FRG-Flat ZF. From Table~\ref{tab:exp1}, we also observe that the WMMSE-type precoders in general outperform the ZF-type precoders. In the forthcoming simulations, we examine the circumstances under which the performance disparity between flat ZF and flat WMMSE becomes more notable.

\begin{figure}[t]
  \centering
  \begin{subfigure}{2.8in}
    \includegraphics[width=\textwidth]{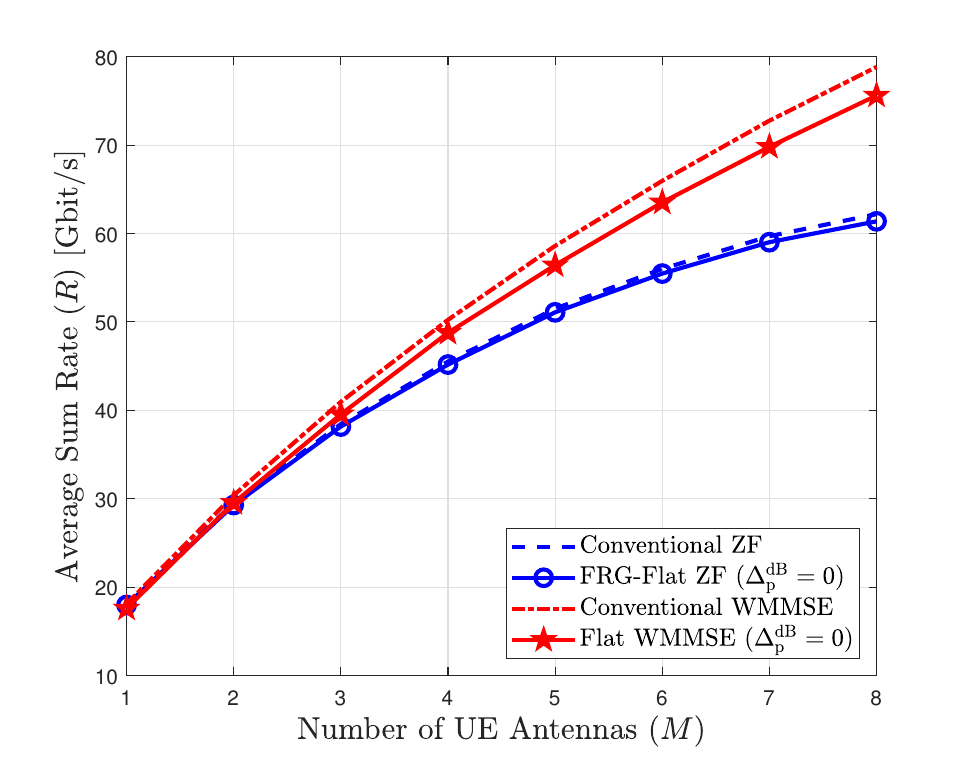}
    \caption{Average sum rate versus $M$.}
    \label{subfig:exp3_1}
  \end{subfigure}\\%
  \begin{subfigure}{2.8in}
    \includegraphics[width=\textwidth]{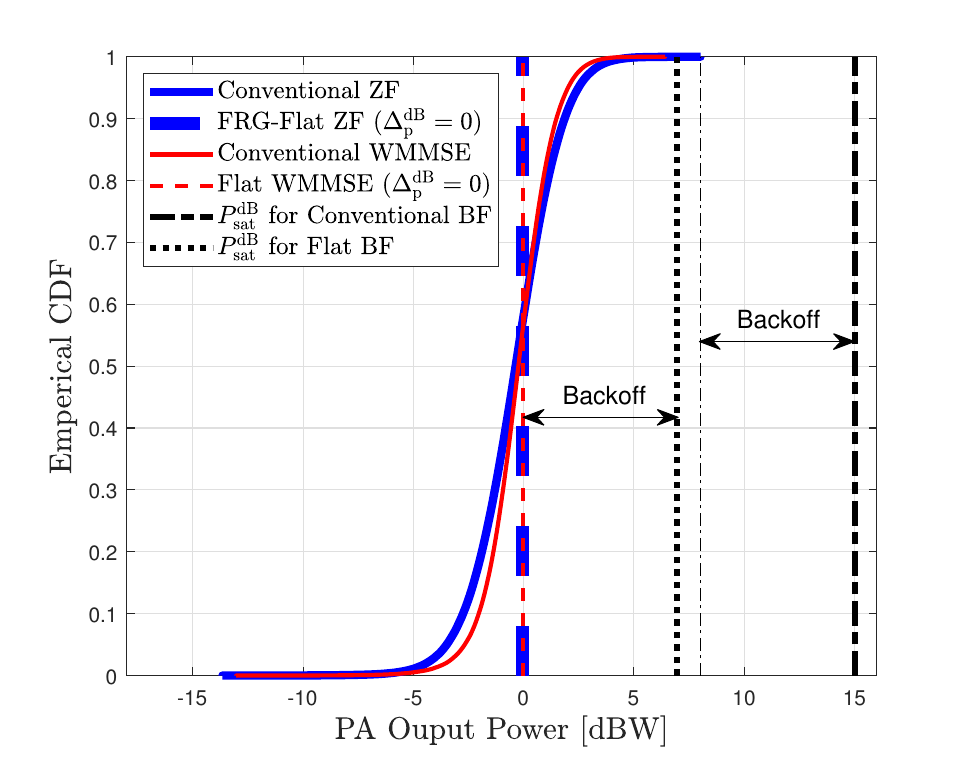}
    \caption{\changeb{PA output power CDF (aggregated over $M=1,...,8$).}}
    \label{subfig:exp3_2}
  \end{subfigure}\\
  \begin{subfigure}{2.8in}
    \includegraphics[width=\textwidth]{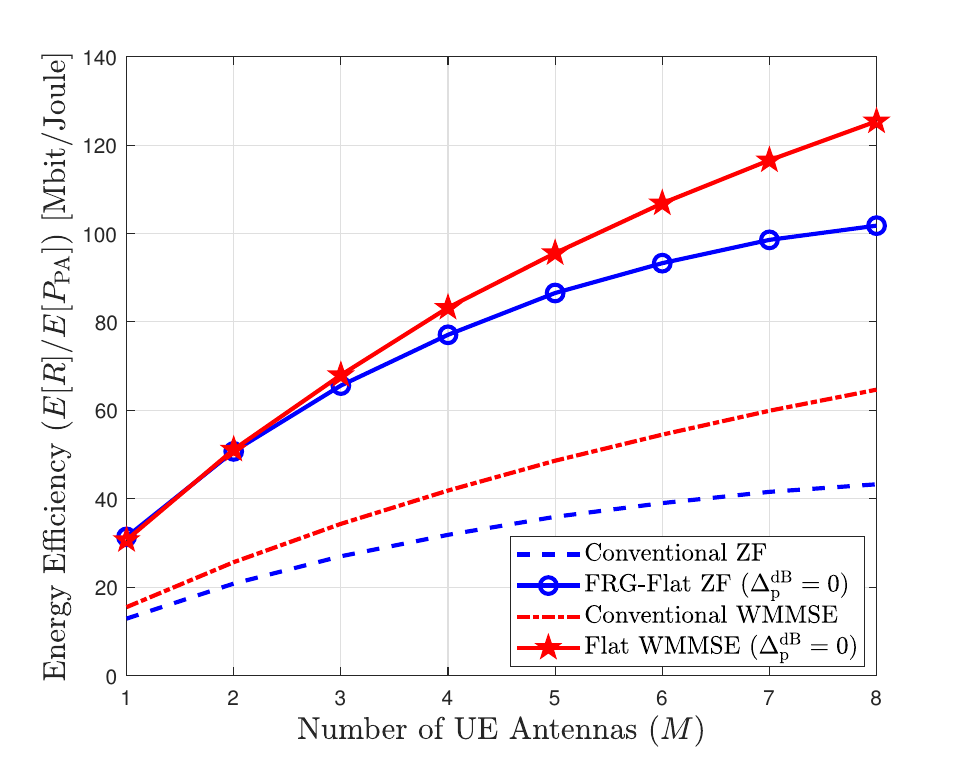}
    \caption{Energy efficiency versus $M$.}
    \label{subfig:exp3_3}
  \end{subfigure}
  \caption{The performance comparison between different methods in a MU-MIMO system with $K=8$ UEs (each equipped with $M$ antennas), $L=M$ layers per UE, and $N=256$ BS antennas.}
  \label{fig:exp3}
\end{figure}
 
For the second experiment, we maintain all parameters identical to those used in the first experiment and solely vary the number of BS antennas $N$ from $8$ to $256$. Subsequently, we illustrate the sum rate performance, the required PA saturation power, and energy efficiency for different methods against varying $N$ in Fig.~\ref{fig:exp2}. From Fig.~\ref{subfig:exp2_1}, it can be observed that imposing flatness results in only a marginal degradation in the sum rate performance. Further, for MU-MISO systems, the performance gap between ZF- and WMMSE-type precoders is negligible, especially if $K\ll N$. For scenarios where $N$ is not significantly larger than $K$, WMMSE precoding outperforms ZF precoding since enforcing the ZF structure in a limited dimension can be more challenging (and even infeasible). Fig.~\ref{subfig:exp2_2} illustrates that embracing the flat beamforming idea results in a substantial reduction in the PA saturation level, leading to a subsequent reduction in PA size, and ultimately contributing to a lower implementation cost of the system. Finally, Fig~\ref{subfig:exp2_3} indicates that whether we consider a ZF structure or a WMMSE structure, flat precoding can significantly enhance the overall energy efficiency of the system as compared to the corresponding conventional methods.
 
In the next experiment, we focus on the MU-MIMO setting where a BS with $256$ antennas serves $K=8$ UEs, each equipped with $M$-antennas. Further, the number of intended layers per UE is set to be $L=M$. Fig.~\ref{subfig:exp3_1}, which plots the average sum rate against $M$, again verifies that imposing flatness, either with ZF- or WMMSE- structure, does not significantly impact the sum rate performance. Moreover, we can see that, in the MU-MIMO systems, the gap between ZF and WMMSE widens as the number of UE antennas increases. This is most likely because, in the adopted ZF precoder, we enforce ZF structure even between different layers intended for a UE. Therefore, the number of ZF constraints scales with the number of UE antennas/intended layers. It should be mentioned that it is conceivable that by using more sophisticated ZF methods such as eigen ZF \cite{Sun2010EZF} or block diagonal ZF \cite{Lu2020DZF}, this gap may be further reduced. However, developing a flat precoding algorithm based on such more sophisticated ZF methods is beyond the scope of this paper. Nevertheless, it is a promising direction for future research. Further, based on Fig.~\ref{subfig:exp3_2} and Fig.~\ref{subfig:exp3_3}, we can see that the same observations can be made for MU-MIMO case as in the MU-MISO case. In particular, Fig.~\ref{subfig:exp3_2} shows that the PA saturation power (and equivalently the PA size) required for flat precoding is much smaller as compared to that for the conventional precoding schemes; and Fig.~\ref{subfig:exp3_3} illustrates that the flat precoding methods are much more energy efficient as compared to conventional precoders.

\begin{figure}[t]
  \centering
  \includegraphics[width=2.8in]{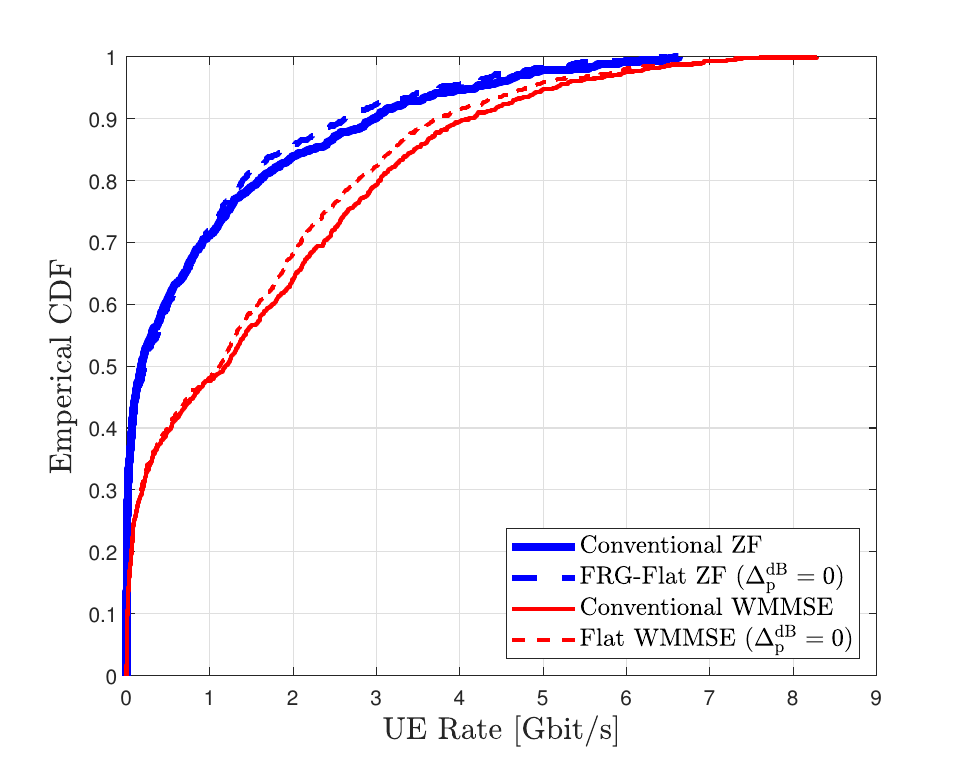}
  \caption{The CDF curve of the UE rate in a MU-MISO system with $K=8$ UEs and $N=16$ BS antennas.}
  \label{fig:exp4}
\end{figure}

Finally, we aim to show that the proposed flat precoding methods can also handle the WSR maximization problem. Here, we set $N=16$ and $M=L=1$. Further, by using the principles in \cite{Mitran2021Weights}, we set the priority weights based on the long-term rate achieved by each UE so far and consider $T = 100$ time frames for calculating the weights.
Based on the rate CDF curve in Fig.~\ref{fig:exp4}, it is again verified that imposing flatness is nearly cost-free from a rate performance perspective. Moreover, the WMMSE method can achieve a higher performance as compared to the ZF method, especially in scenarios like this simulation where the number of BS antenna $N$ is not much larger than number of the UEs $K$.

\section{Conclusion}
\label{sec:conc} 

 This paper introduces a novel energy-efficiency concept called flat precoding, which aims to control power variations across antennas. Multiple algorithms are developed to design WMMSE- or ZF-type precoders with a controllable flatness feature. The results demonstrate that, given the current PA technology, complete flat precoding approaches offer the optimal balance between spectral efficiency and energy efficiency.  Overall, our findings highlight the great potential of flat beamforming to enhance energy efficiency and to reduce implementation costs in MIMO systems.

\changeb{
\section*{Acknowledgment}
We would like to acknowledge and appreciate the anonymous reviewers for their insightful comments and suggestions, which significantly improved the clarity and rigor of this paper.
}
\bibliographystyle{IEEEtran}
\bibliography{IEEEabrv,referenceF}

\end{document}